\documentclass[aps,prb,twocolumn,amsmath,superscriptaddress,nofootinbib,amssymb,10pt]{revtex4-1}

\makeatletter

\newcommand{\Rmnum}[1]{\expandafter\@slowromancap\romannumeral #1@}
\makeatother

\usepackage{amsmath}
\usepackage{tabularx,graphicx}
\usepackage{epstopdf}
\usepackage{graphicx}
\usepackage{latexsym}
\usepackage{amssymb}
\usepackage{amsmath}
\usepackage{color}
\usepackage{psfrag}
\usepackage{bbm}
\usepackage{bm}
\usepackage{titlesec}
\usepackage{dsfont}
\usepackage{feynmp}
\usepackage{slashed}

\newcommand{\mc}[1]{\mathcal{#1}}

\newcommand{\beq}{\begin{eqnarray}}
\newcommand{\eeq}{\end{eqnarray}}

\newcommand{\la}{\langle}
\newcommand{\ra}{\rangle}

\newcommand{\bsp}{\begin{split}}
\newcommand{\esp}{\end{split}}

\begin{document}
\title{Dimensional decoupling at continuous quantum critical Mott transitions}
\author{Liujun Zou}
\affiliation{Department of Physics, Harvard University, Cambridge, MA 02138, USA}
\author{T. Senthil}
\affiliation{Department of Physics, Massachusetts Institute of Technology, Cambridge, MA 02139, USA}
\date{\today}

\begin{abstract}
For continuous Mott metal-insulator transitions in layered two dimensional systems, we demonstrate the phenomenon of dimensional decoupling: the system behaves as a three-dimensional metal in the Fermi liquid side but as a stack of decoupled two-dimensional layers in the Mott insulator. We show that the dimensional decoupling happens at the Mott quantum critical point itself. We derive the temperature dependence of the interlayer electric conductivity in various crossover regimes near such a continuous Mott transition, and discuss  experimental implications.
\end{abstract}

\maketitle

\section{Introduction} \label{secintro}

Despite its great success in describing many metals,  Fermi liquid theory fails to characterize many strongly correlated metals.  Examples of ``non-Fermi liquid" metals found experimentally include heavy fermion compounds, the cuprates,  and some organic salts\cite{LoehneysenPietrusPortischEtAl1994,CustersGegenwartWilhelmEtAl2003, Lee2009,KurosakiShimizuMiyagawaEtAl2005, FurukawaMiyagawaTaniguchiEtAl2015,OikeEtAl2015,OikeSuzukiTaniguchiEtAl2016}.

The central conceptual building block in Fermi liquid theory is the existence of  long-lived electronic quasiparticles near a sharply defined Fermi surface in momentum space. The electron spectral function $\mc{A}(\vec k,\omega)$ near the Fermi surface
takes the form
\beq
\mc{A}(\vec k,\omega)=Z\delta(\omega-\epsilon(\vec k))
\eeq
where $Z$, the quasiparticle residue, measures the overlap between the wave function of a quasiparticle and that of the original electron, and $\epsilon(\vec k)$ is the (gapless) dispersion of quasiparticles.
Metals in which this building block breaks down will show  non-Fermi liquid properties in a number of experimental probes.

Our concern in this paper is on non-Fermi liquid metals in layered quasi-two dimensional systems near Mott metal-insulator (and other closely related) phase transitions. Specifically we will focus on metals at or near continuous ({\em i.e} quantum critical) Mott transitions in such systems. Previous work has  demonstrated, for  a single isolated two dimensional ($2d$) layer, the possibility of a second order quantum phase transition from a Fermi liquid metal to a quantum spin liquid Mott insulator with a Fermi surface of charge neutral spin-$1/2$ fermions\cite{Senthil2008a}. These studies were motivated by the phenomenology of the quasi-$2d$ triangular lattice organic materials and the cuprate metals\cite{Lee2009,KurosakiShimizuMiyagawaEtAl2005}, and they have obtained support from numerical studies\cite{MishmashGonzalezMelkoEtAl2015}.
Here we study the effect of weak interlayer coupling on the fate of such continuous Mott transitions in the physical three dimensional material.

In the specific context of the cuprate metals it has long been appreciated  that the combination of their  layered quasi-two dimensional structure and their possible non-Fermi liquid properties could lead to peculiar interlayer transport. Specifically interlayer transport was argued to be related to the single particle spectrum in such layered materials, and hence probe very different physics from intralayer transport\cite{Andersonothers1997}.
Interlayer transport is thus a very useful spectroscopic probe of a correlated quasi-two dimensional metal, particularly in situations where other direct probes like photoemission or tunneling is not feasible\cite{IoffeMillis1999}.

\begin{figure}[h!]
  \centering
  \includegraphics[width=0.36\textwidth]{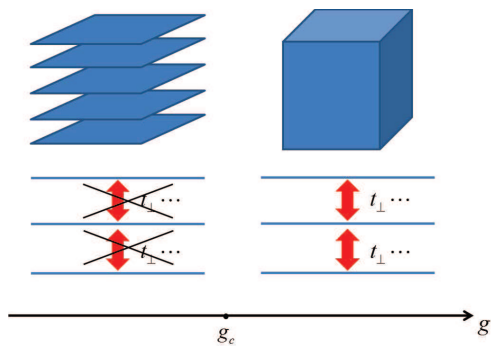}\\
  \caption{Dimensional decoupling across the phase transition. On the side with $g>g_c$ we have a 3D Fermi liquid, and at the quantum critical point and on the other side we have a state that behaves as a stack of many decoupled layers at low energies.}\label{Dimensional decoupling diagram}
\end{figure}

We  show that, in the spin liquid Mott insulating phase, different layers decouple from each other so that the system behaves as a stack of two dimensional layers. We call this `dimensional decoupling'. In contrast, in the Fermi liquid, different layers recouple to form a coherent three dimensional Fermi surface (see Fig. \ref{Dimensional decoupling diagram}). The transition is thus between a three dimensional Fermi liquid and a dimensionally decoupled stack of two dimensional insulators. We show that this phase transition is continuous, and further that the dimensional decoupling happens already at the quantum critical point. In other words, universal critical properties are correctly obtained from a purely two dimensional theory though the metallic phase on one side is (at low energies) three dimensional. We discuss the implications for physical properties as the transition is approached from the metallic side.  In particular we show that the interlayer conductivity as a function of decreasing temperature in the nearly critical metal has a `coherence' peak at a crossover scale determined by the distance from the quantum critical point, and determine the detailed universal temperature dependence in various regimes.

Very recent experiments have studied interlayer transport in the doped triangular lattice $\kappa-ET$ organic metals\cite{OikeSuzukiTaniguchiEtAl2016}, and our results should be a useful guide to their interpretation. Ref. \onlinecite{OikeSuzukiTaniguchiEtAl2016} found the intralayer transport appears to be non-Fermi liquid like at ambient pressure and becomes Fermi liquid like under high enough pressure. As the temperature is decreased in the non-Fermi liquid regime, the interlayer resistivity first increases when the temperature is high. But it starts to decrease when the temperature is low enough, so it has a peak as temperature changes. This `interlayer coherence' peak becomes broadened and is shifted to higher temperatures as the pressure becomes higher. Further this peak seems to occur in a regime where the intralayer transport is non-Fermi liquid like. It is thus timely to study issues related to interlayer coupling near continuous Mott transitions.

In the cuprate context  previous theoretical work on doped Mott insulators identified, within a slave boson framework,  an ``Incoherent Fermi Liquid" (IFL) regime that has some phenomenological appeal as a description  of the strange metal normal state\cite{SenthilLee2009}. Our results describe the interlayer transport of this Incoherent Fermi Liquid. Our results also carry over straightforwardly to  Kondo breakdown transitions (of the kind studied in Refs. \onlinecite{SenthilVojtaSachdev2004}) in Kondo lattice models in layered  systems.

We emphasize that dimensional decoupling is not guaranteed to describe {\em all} two dimensional non-Fermi liquids.  To illustrate this we consider a class of non-Fermi liquids that develop at metallic quantum critical points associated with onset of broken symmetry. A good and topical example is the onset of Ising nematic ordering from a symmetry preserving metal.  In Appendix \ref{appisingnematics} we show, within the existing theory of this transition, that at the quantum critical point   there is no dimensional decoupling. We provide arguments that this is likely the case at all Landauesque quantum critical points driven by broken symmetry order parameter fluctuations (coupled to the metallic electrons).

In contrast continuous Mott transitions or the Kondo breakdown transitions are driven by electronic structure fluctuations which cannot be captured through Landau order parameters.  On approaching this kind of quantum critical point from the metallic Fermi liquid side the quasiparticle residue $Z$ is expected to vanish continuously.  Right at the
critical point the quasiparticle is thus destroyed everywhere. Despite this,
Ref. \onlinecite{Senthil2008} argued that the quantum critical point is characterized by a sharply defined Fermi surface (dubbed a ``critical Fermi surface").   Concrete examples which illustrate the general arguments of Ref. \onlinecite{Senthil2008} are in Refs. \onlinecite{Senthil2008a,SenthilVojtaSachdev2004,NandkishoreMetlitskiSenthil2012} based on slave-particle gauge theories.

Clearly such phase transitions driven by electronic structure fluctuations require a different conceptual framework from more conventional order parameter driven ones. The corresponding quantum critical phenomenology will also be very different.
Our results add to the growing list of distinctions between  these two classes of metallic quantum critical phenomena.

To address the possibility of dimensional decoupling in these systems, it is necessary to  examine the effects of all possible interlayer interactions. The simplest amongst these is electron tunneling between different layers. Other potentially important couplings is that between slowly fluctuating bosonic order parameters obtained as bilinears made out of the underlying electrons.  The phenomenon of dimensional decoupling requires that,  in the renormalization group (RG) sense, none of the possible interlayer interactions  is relevant at
the decoupled fixed point with no interlayer coupling. For the order parameter driven phase transitions, the coupling between order parameter fluctuations in different layers becomes relevant and destabilizes the decoupled fixed point.  However this does not happen at the continuous Mott transition.


In passing, we note that our starting point is that the coupling between different layers is weak at the lattice level, so it is legitimate to first consider the system as a stack of decoupled layers and then study the effects of the interlayer couplings. In contrast, other works, such as Ref. \onlinecite{PodolskyParamekantiKimEtAl2009}, study the situation where the interlayer interactions are as strong as intralayer interactions, and in this case considering a 3D system from the beginning is more appropriate.

The rest of this paper is organized as follows. We begin with a brief review of some general (semi-) quantitative aspects of a critical Fermi surface and continuous Mott transitions in Sec. \ref{prelim}. Before diving into the more complicated case of continuous Mott transitions, in Sec. \ref{secom} we will warm up by illustrating the phenomenon of dimensional decoupling in a simpler context: a quantum phase transition between a Fermi liquid and an orthogonal metal. We will first consider a single 2D layer and review the nature of the orthogonal metal state and the phase transition in 2D, then we consider a stack of many such 2D layers and examine the effects of interlayer interactions. Following the same strategy, we will go into the case of continuous Mott transitions and demonstrate dimensional decoupling thereof in Sec. \ref{seccmt}. After this, in Sec. \ref{secconductivity} we will calculate the interlayer electric conductivity induced by electron tunneling between different layers in the QC regimes of the various cases of interests. Because of the rich crossover structure predicted by the theory of continuous Mott transitions, in order to identify some of its experimental signatures, in Sec. \ref{seccrossover} we study the interlayer electric conductivity as the system crosses over from its QC regimes to the Fermi liquid regime. Finally, we conclude with some discussions on experiments in Sec. \ref{secdiscussion}.

\section{Preliminaries}\label{prelim}
In this section we collect together some previous results on critical Fermi surfaces and continuous Mott transitions that will be used extensively in the rest of this paper.

\subsection{Critical Fermi surface}

A continuous Mott transition from a Fermi liquid metal requires a sudden death of the the metallic Fermi surface.  The transition to the Mott insulator occurs without the vanishing of the free carrier density, and hence the Fermi surface cannot simply shrink to zero. Thus continuous Mott transitions necessarily involve the death of an entire  Fermi surface of some fixed size. Ref. \onlinecite{Senthil2008} argued that at the corresponding quantum
critical point there will be a sharp critical Fermi surface but without well-defined Landau quasiparticles.  To discuss interlayer coupling effects in different systems in a unified manner, we will need some very general scaling properties of the electron spectral function at such a critical Fermi surface which we now summarize.

 Suppose one can access such a quantum critical point by tuning a parameter $g$.  It is natural that at temperature $T$, the electron spectral function around the critical Fermi surface near such a critical point has the scaling form:
\beq \label{cfsscaling}
\mc{A}(\vec K,\omega;g,T)\sim\frac{c_0}{|\omega|^{\frac{\alpha}{z}}}F\left(\frac{\omega}{T}, \frac{c_1k_\parallel}{T^{\frac{1}{z}}},k_\parallel\xi\right)
\eeq
where $\alpha$ and $z$ are two universal exponents, and $F$ is a universal function. $k_\parallel$ is the deviation of momentum $\vec K$ from the Fermi surface. Constants $c_0$ and $c_1$ are non-universal. $\xi$ is the correlation length of the system at the same $g$ but at zero temperature, and it is a measure of the deviation of $g$ from the quantum critical point $g_c$ since $\xi^{-1}\sim|g-g_c|^\nu$, with a universal zero temperature correlation length exponent $\nu$. This scaling form of the spectral function applies to each patch of the Fermi surface, and, in general, $\alpha$, $z$, $\nu$, $c_0$ and $c_1$ can all depend on the position of the patch on the Fermi surface where (\ref{cfsscaling}) is applied. We will take the convention that the Fermi liquid is on the $g>g_c$ side through out this paper.

According to (\ref{cfsscaling}), the number of fermions with a certain momentum $\vec k$ is
\beq
n(\vec k)=\int_{-\infty}^0d\omega\mc{A}_c(\vec k,\omega)\sim|k_\parallel|^{z-\alpha}
\eeq
Because the fermion number is upper bounded, we must have
\beq \label{zalphabound}
z\geqslant\alpha
\eeq

Again, we notice that (\ref{cfsscaling}) and (\ref{zalphabound}) apply to each patch of the Fermi surface, and in general the exponents can depend on the position of patch. The same scaling considerations apply to all systems with a critical Fermi surface.

\subsection{Theory of a continuous $2d$ Mott transition}

We consider two types of continuous Mott transitions: chemical potential tuned   and bandwidth tuned, for which in a single $2d$ layer there are well developed theories.  Theoretically, it is expected that these Mott transitions can be realized by the Hubbard-type Hamiltonian on a triangular lattice
\beq \label{Hubbard}
\begin{split}
H&=-t\sum_{ij,\sigma}(c_{i\sigma}^\dag c_{j\sigma}+{\rm h.c.})\\
&\qquad\qquad\quad
+U\sum_in_{i\uparrow}n_{i\downarrow}-\mu\sum_{i,\sigma}n_{i\sigma}
\end{split}
\eeq
where $c_{i\sigma}$ annihilates an electron with spin $\sigma$ at site $i$, and $n_{i\sigma}=c_{i\sigma}^\dag c_{i\sigma}$ is the number operator of the fermion. There is numerical evidence for a regime described by a Mott insulator with a spinon Fermi surface \cite{Motrunich2005,LeeLee2005,ShengMotrunichFisher2009, YangLaeuchliMilaEtAl2010}. Experimentally the organic compound $\kappa-(ET)_2Cu_2(CN)_3$, which is believed to be well described by a one-band Hubbard model on a triangular lattice, also exhibits signatures of these transitions \cite{KurosakiShimizuMiyagawaEtAl2005,OikeEtAl2015, FurukawaMiyagawaTaniguchiEtAl2015}.

The transition is conveniently accessed by formally writing the electron operator $c_{i\sigma}$ as
\beq
c_{i\sigma}=f_{i\sigma}\cdot b_i
\eeq
where the fermionic spinon $f$ carries spin-$\frac{1}{2}$, and the boson carries unit physical charge. The physical electron operator is invariant under a local $U(1)$ gauge transformation $f_{i\sigma}\rightarrow f_{i\sigma}e^{i\phi_i}$ and $b_i\rightarrow b_ie^{-i\phi_i}$, which leads to an emergent $U(1)$ gauge field at low energies. Consider a situation where spinons form a Fermi surface. If the boson $b$ condenses, the gauge field is Higgsed out and we get a Fermi liquid of the original electron. If $b$ is gapped, we get a spin liquid with a spinon Fermi surface coupled to a $U(1)$ gauge field. The longitudinal component of the gauge field is screened by the spinon Fermi surface, but the transverse components remain strongly interacting with the spinons. The Mott transition from the Fermi liquid is driven by losing the condensate of $b$.


For both the chemical potential and bandwidth tuned transitions, Ref. \onlinecite{Senthil2008a} showed that the quantum critical fluctuations of $b$ (at $T = 0$) are dynamically decoupled from the spinon-gauge system. However, the dynamics of the latter is affected by the criticality of $b$. This leads to a tractable theory of these continuous Mott transitions and
a number of universal physical properties have been computed. An interesting feature shared by both transitions is that the crossover from the quantum critical metal to the Landau Fermi liquid on the metallic side occurs in two stages. The charge sector crosses over at an energy scale parametrically larger than the spin sector. At intermediate energies a non-Fermi liquid metallic regime is reached which is distinct from the quantum critical non-Fermi liquid.

We will study the effects of interlayer coupling on these transitions, and determine the nature of interlayer transport both in the quantum critical non-Fermi liquid and in the intermediate energy non-Fermi liquid that exists before the emergence of the fully coherent Landau Fermi liquid.

Notice in principle this $U(1)$ gauge field should be taken to be compact, but as pointed out in Ref. \onlinecite{Lee2008}, the effect of instantons of this compact $U(1)$ gauge field will be suppressed by the spinon Fermi surface, so it is adequate to just consider a noncompact $U(1)$ gauge field. It is also sufficient for our purposes to  treat this non-compact $U(1)$ gauge field under a random-phase-approximation (RPA), which can be formally justified by a controlled expansion in its leading order \cite{MrossMcGreevyLiuEtAl2010}.

\section{Warm-up: Dimensional decoupling in a quantum phase transition between a Fermi liquid and an orthogonal metal}
 \label{secom}
To examine effects of interlayer coupling  at continuous Mott transitions we need to confront the full theory of the the critical $b$ fluctuations and the spinon-gauge system. In this section we warm up to this task by considering a different problem which has some of the same ingredients. Rather than studying the transition from a Fermi liquid to a Mott insulator, we study the transition to a different phase dubbed the orthogonal metal. We demonstrate the phenomenon of dimensional decoupling at this quantum phase transition.

An orthogonal metal is a state where the electron is fractionalized into a fermion $f_{i\sigma}$ that carries both the charge and spin of the electron and  a discrete degree of freedom $s_i$ that is gapped \cite{NandkishoreMetlitskiSenthil2012}.  This fractinonalization is accompanied by  a deconfined discrete gauge field to which both $f_{i\sigma}$ and $s_i$ are coupled. The $f_{i\sigma}$ forms a Fermi surface, and the system has metallic charge/spin transport. However the gapless $f_{i\sigma}$ have zero overlap with the physical electrons (they are orthogonal) . Thus despite the metallic charge transport single partilce-probes like tunneling or photoemission will see insulating behavior. The orthogonal metal is the simplest non-Fermi liquid in spatial dimension $d \geq 2$, and its universal properties are easily computed. Some lattice models that can realize this state are provided in Ref. \onlinecite{NandkishoreMetlitskiSenthil2012}.

Formally we write the electron operator $c_{i\sigma}$ as
\beq \label{fractionalizationom}
c_{i\sigma}=f_{i\sigma}\cdot s_i
\eeq
Assuming the fermions $f$ are in {\it their} Fermi liquid phase, if $s$ condenses, the resulting state is a Fermi liquid of the original electrons. However, if $s$ is gapped, we get an orthogonal metal, whose nature will be determined by the nature of the discrete variable $s$.


In passing, we remark that all systems discussed in this paper have a Fermi surface, either of the physical electrons or of some emergent fractionalized fermions. It is well-known that a Fermi surface may potentially suffer from an instability towards Cooper pairing, but because this instability occurs only at very low temperatures, we will assume our systems are in a regime free of pairing instability through out this paper. We also notice that in some systems with a Fermi surface coupled to a gauge field, the pairing instability is suppressed\cite{MetlitskiMrossSachdevEtAl2015}.

Now consider a single 2D layer first. One can drive a transition from a Fermi liquid of electrons, a condensate of $s$, to an orthogonal metal by destroying the condensate. As discussed in Ref. \onlinecite{NandkishoreMetlitskiSenthil2012}, in 2D the transition to a $Z_2$ orthogonal metal in a lattice needs fine-tuning of the parameters in the Hamiltonian, and the simplest example where a generic second-order phase transition can occur in a 2D lattice is between an electronic Fermi liquid and a $Z_4$ orthogonal metal. The critical theory for such a transition is described by the following Lagrangian
\beq
\begin{split}
\mc{L}
=&b^*\partial_\tau b+\frac{1}{2m_b}|\nabla b|^2+t|b|^2+\frac{u}{4}|b|^4\\
&+\frac{v}{4!}\left[b^4+(b^*)^4\right]
\end{split}
\eeq
where the $b$ transforms as $b\rightarrow ib$ under the $Z_4$ transformation. Notice this is not a symmetry-breaking phase transition because $b$ itself is not gauge invariant. Instead, this is a transition associated with electron fractionalization.

\begin{figure}[h!]
  \centering
  \includegraphics[width=0.3\textwidth]{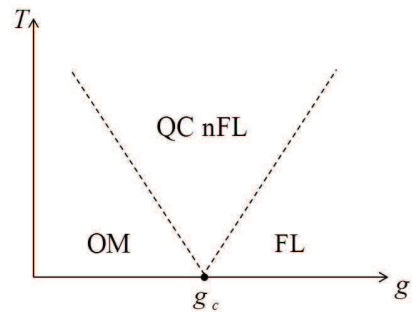}\\
  \caption{The schematic phase diagram and crossover structure of an orthogonal metal transition. When $g>g_c$ we have a Fermi liquid (FL), while we have an orthogonal metal (OM) when $g<g_c$. At finite temperature, there is a quantum critical regime (QC) where the system is non-Fermi liquid like.}\label{OM diagram}
\end{figure}

At the critical point,  the quasiparticle residue of the electron vanishes and the excitations are incoherent. The spectral function of the $Z_4$ spins at the critical point is $\mc{A}_{s}(\vec q,\Omega)\sim\delta(\Omega-\frac{\vec q^2}{2m_b})$. Since the spinons are in their Fermi liquid phase, they have spectral function $\mc{A}_f(\vec k,\omega)\sim\delta(\omega-v_Fk_\parallel)$. Convolving them we get the electron spectral function at the quantum critical point:
\beq \label{elecspectom}
\begin{split}
\mc{A}_c(\vec k,\omega)
=&\int_{\vec q}\int_0^\omega d\Omega\mc{A}_{s}(\vec q,\Omega)\mc{A}_f(\vec k-\vec q,\omega-\Omega)\\
\sim&\left(\omega-\frac{k_\parallel^2}{2m_b}\right)^{\frac{1}{2}}\theta\left(\omega-\frac{k_\parallel^2}{2m_b}\right)
\end{split}
\eeq
This has the form of (\ref{cfsscaling}), and  we see that this orthogonal metal transition has a critical Fermi surface with $\alpha=-1$ and $z=2$. As we can also see, (\ref{zalphabound}) is indeed satisfied. The large and negative value of $\alpha$ means in the quantum critical regime (QC) above the quantum critical point, the system is highly non-Fermi liquid like. If $g\neq g_c$, starting from the QC regime, as the temperature is decreased, the system crosses over to a Fermi liquid metal or an orthogonal metal, depending on the relative magnitude of $g$ compared to $g_c$ (see Fig. {\ref{OM diagram}}).

Now consider a stack of such 2D systems, with each layer going through a phase transition between a Fermi liquid and a $Z_4$ orthogonal metal. When the interlayer interactions are absent, in terms of RG, the critical point is described by a fixed point where each layer corresponds to an individual fixed point and all these fixed points are decoupled. We will call such a fixed point a ``decoupled fixed point". Let us examine the effect of all possible interlayer interactions on this decoupled fixed point. Note that the decoupled fixed point has separate conservation of physical electric charge in different layers corresponding to independent global $U(1)$ symmetry rotations in each layer. In writing (\ref{fractionalizationom}), we have introduced a $Z_4$ gauge redundancy on each layer, so the interlayer interactions should be invariant under a local $Z_4$ gauge transformation within {\it each} layer. The most obvious physical coupling is simply electron tunneling between different layers - this breaks the infinite number of $U(1)$ symmetries associated with conservation of electric charge separately in each layer to a single common global $U(1)$. We will first however focus on interlayer couplings that preserve this infinite $U(1)$ symmetry.

The most important such interactions consistent with gauge invariance and global symmetries are the coupling between energy densities of different layers, which is of the form
\beq \label{interlayercoupling1}
\delta\mc{L}_1=\sum_{\alpha}\int d\tau d^2xg_{1\alpha\beta}|b_\alpha|^2|b_\beta|^2
\eeq
and the coupling between the energy density of one layer and the collective excitation around the spinon Fermi surface of the other layer, which is of the form
\beq \label{interlayercoupling2}
\delta\mc{L}_2=\sum_{\alpha\beta,\sigma}\int d\tau d^2xg_{2\alpha\beta}|b_\alpha|^2f_{\beta\sigma}^\dag f_{\beta\sigma}
\eeq
Here we use $\alpha$ and $\beta$ to index the layers.

Perturbing the decoupled fixed point with $\delta\mc{L}_1$, we get its RG equations
\beq \label{oneloopRG}
\frac{dg_{1\alpha\beta}}{dl}=-\mc{C}g_{1\alpha\beta}^2
\eeq
with a constant $\mc{C}>0$ (see Appendix \ref{apponeloop}). From this RG equation, we see $\delta\mc{L}_1$ is (marginally) irrelevant.

On the other hand, to consider the effect of $\delta\mc{L}_2$ on the decoupled fixed point, it is convenient to integrate out the degrees of freedom from the spinons, and the most relevant interaction generated from this procedure has the following Landau damping form
\beq \label{fermionsintegratedout}
\sum_{\alpha}g_{2\alpha\beta}'\int_{\omega,\vec q}\Pi(\vec q,\omega)\cdot|b_\alpha|^2(\vec q,\omega)\cdot|b_\beta|^2(\vec q,\omega)
\eeq
with
\beq \label{densitycorrelator}
\Pi(\vec q,\omega)\sim\frac{|\omega|}{|\vec q|}
\eeq
for small frequencies and momenta. Because the decoupled fixed point has dynamical exponent $z=2$, the dependence of $\Pi$ on the frequency and momentum $\frac{|\omega|}{|\vec q|}\sim q$ makes this interaction irrelevant.

As for other interactions, one should in principle consider the couplings between charge densities, spin densities, charge currents and spin currents, and the most relevant ones of these couplings in this case are of the form of four-fermion interactions of the spinon $f$. It is well-known that in the presence of a Fermi surface, most of the four-fermion interactions are strongly irrelevant, except for the forward scattering and the BCS scattering. The former is marginal and will not modify the physical properties of the system qualitatively, and they can be described by a set of Landau parameters, while the latter is marginally irrelevant and can induce pairing instability at very low temperatures\cite{Shankar1994,Polchinski2008,Shankar2005}. In our case, most of these interlayer four-fermion interactions are also strongly irrelevant due to the kinematic constraint of the Fermi surface, and the analog of forward scattering will renormalize the in-plane Landau parameters, but they will not give rise to any qualitative change of the physics. The analog of BCS scattering can in principle induce interlayer pairing at very low temperatures, but as declared before, we will ignore it.

We now return to the important effect of interlayer electron tunneling. To discuss these we will use a scaling argument that is somewhat different from the one above.

Consider a general action for the interlayer electron hopping. At low energies it is appropriate to work with electronic modes  near the critical Fermi surface.  The hopping term can then be written as
\beq
\begin{split}
\delta S_{\rm tunneling}
=&-\int d\omega dk_\parallel d\theta\sum_{\gamma\delta} t_{\gamma\delta}(\theta)\cdot\\
&\quad\left( c_\gamma^\dag(k_\parallel,\theta,\omega)c_{\delta}(k_\parallel,\theta,\omega) +{\rm h.c.}\right)
\end{split}
\eeq
Now for any critical Fermi surface consider a scaling transformation that renormalizes toward the Fermi surface. We let $k_\parallel \rightarrow k_\parallel' = k_\parallel s$, and $\omega \rightarrow \omega' = \omega s^z$.   The two point correlation function of
$c(k_\parallel,\theta,\omega)$ satisfies
\beq
\begin{split}
\langle c(k_{\parallel 1},\theta,\omega_1)c^\dagger(k_{\parallel 2},\theta,&\omega_2) \rangle=\\
 & \delta(k_{\parallel 1} - k_{\parallel 2}) \delta(\omega_1 - \omega_2) G(k_{\parallel 1}, \theta, \omega_1) \nonumber
\end{split}
\eeq
If the electron spectral function for a 2D layer has the scaling form (\ref{cfsscaling}), then the the electron operator transforms as  $c(k_\parallel,\theta,\omega) \rightarrow c'(k_\parallel', \theta, \omega') = s^{-\frac{\alpha+z+1}{2}}c(k_\parallel,\theta,\omega)$. It follows that the scaling of the hopping parameter is $t'(\theta)=t(\theta) s^\alpha$.  If $\alpha<0$, we expect that the interlayer electron hopping is irrelevant.

As discussed above the electron spectral function for a 2D layer at the critical point for the Fermi liquid to orthogonal metal transition satisfies the scaling form of Eqn.  \ref{cfsscaling} with $\alpha = -1$ and $z = 2$., and thus the inter electron hopping indeed scales to zero at low energy.

We point out a caveat in this scaling analysis. In the example of orthogonal metal transition discussed in this section (and the chemical potential tuned Mott transition discussed later), the boson sector has dynamical exponent $z_b=2$ while the fermion sector has dynamical exponent $z_f=1$, and the electron Green's function has dynamical exponent $z_c=2$. So when we do a scaling analysis, how should we scale space and time? Notice the above dynamical exponents indicate the important dynamical regions are $\omega\sim k^2$ and $\omega\sim k$, and the former governs the important electron dynamics. Moreover, the former is a slower regime compared to the latter. Therefore, to consider the lowest energy physics which is also pertinent to the electrons, we choose $z=2$ in the above scaling analysis.

Therefore, none of the interlayer interactions is relevant at the decoupled fixed point, and different layers do decouple at the quantum critical point. If we go into the orthogonal metal side, since $b$ is gapped there, the above interactions that involve $b$ will be more irrelevant and the other interactions stay as irrelevant as they are at the critical point. So as long as the system leaves the Fermi liquid phase, it behaves as a stack of many decoupled layers, as shown in Fig. \ref{Dimensional decoupling diagram}. This is our simplest example of dimensional decoupling.

Before moving on, we emphasize the important role of the emergent gauge invariance in obtaining dimensional decoupling. Gauge invariance strongly constrains the possible form of the interlayer interactions, and in the absence of this constraint there will be relevant interactions in general (see Appendix \ref{appisingnematics}).

\section{Dimensional decoupling in continuous Mott transitions} \label{seccmt}

Warmed up with the example of the orthogonal metal transition, now we are ready to deal with the more complicated problem of continuous Mott transitions.

\subsection{Chemical potential tuned continuous Mott transition}

If the system is not at half-filling and the transition is accessed by changing the electron doping, or equivalently, by tuning the chemical potential, the low-energy effective Lagrangian of this transition is
\beq \label{eftc}
\mc{L}=\mc{L}_b+\mc{L}_f+\mc{L}_{gauge}
\eeq
with
\beq
\begin{split}
&\mc{L}_b=\bar b\left[\partial_\tau-ia_0-\mu-\frac{(\vec \nabla-i\vec a)^2}{2m_b}\right]b+V(|b|^2)\\
&\mc{L}_f=\bar f\left[\partial_\tau+ia_0-\frac{(\vec \nabla+i\vec a)^2}{2m_f}-\mu_f\right]f\\
&\mc{L}_{gauge}=\frac{1}{4e^2}(\epsilon^{\mu\nu\lambda}\partial_\nu a_\lambda)^2
\end{split}
\eeq
The potential $V(|b|^2)$ can be taken to be of the conventional form $r|b|^2+g|b|^4$. Notice we did not include a direct interaction between bosons and spinons, since this can be shown to be irrelevant \cite{Senthil2008a}. This model has been extensively studied. Closely related models appear in theories of the cuprates\cite{LeeNagaosa1992} and of the Kondo breakdown phenomenon in Kondo lattice systems\cite{SenthilVojtaSachdev2004}.

\begin{figure}[h!]
  \centering
  \includegraphics[width=0.3\textwidth]{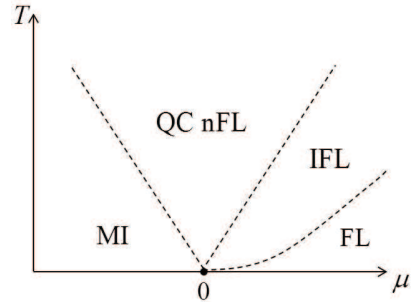}\\
  \caption{Phase diagram and crossover structure of a chemical potential tuned Mott transition. The critical point corresponds to $\mu=0$ and $T=0$. The $\mu<0$ side corresponds to a spin liquid Mott insulator (MI), and the $\mu>0$ side corresponds to a Fermi liquid metal (FL). The quantum critical regime is highly non-Fermi liquid like (QC nFL). In crossing over to the Fermi liquid, the system has to first go through an intermediate incoherent Fermi liquid regime (IFL).}\label{CMT-mu diagram}
\end{figure}

There are several interesting features associated with this transition \cite{Senthil2008a}. First of all, it is shown that at the quantum critical point the bosons are dynamically decoupled from the spinon-gauge-field sector. Therefore, the transition is in the universality class of a dilute (nonrelativistic) bose gas. Using this, one can calculate the electron spectral function at the critical point and show there is indeed a critical Fermi surface, and $\mc{A}(\vec K,\omega)$ in this case has the same form as (\ref{elecspectom}). Since the details of this calculation were not given in Ref. \onlinecite{Senthil2008a}, we present it in Appendix \ref{appelecspecfnchemical}. As shown in Fig. \ref{CMT-mu diagram}, there is a QC regime above the zero temperature quantum critical point, where the system is strikingly non-Fermi liquid like, just as the QC regime of the $Z_4$ orthogonal metal transition.

Moreover, when the system crosses over out from the QC regime to the Fermi liquid, the system has to first go through an intermediate regime. In particular, we can choose the boson phase stiffness $\rho_s$ as the characteristic energy scale on the Fermi liquid side. The system is in its quantum critical (QC) regime if $T\gg\rho_s$. If we decrease the temperature from QC so that $T\ll\rho_s$, the bosons appear to condense. However, the spinon-gauge-field sector has not yet felt the Higgs effect and it behaves as if it is still in its own quantum critical regime. The system in this regime is also a non-Fermi liquid, and it is dubbed ``incoherent Fermi liquid" (IFL)\cite{SenthilLee2009}. When the temperature is further decreased so that $T\ll\rho_s^{\frac{3}{2}}$, the Higgs effect is manifested and the spinon-gauge-field sector also crosses over out of its quantum critical regime, and the system appears as a Fermi liquid. These results can be shown, for example, by calculating the electron spectral function in various regimes.

On the Fermi liquid side, the propagator of the transverse components of the gauge field in Coulomb gauge under RPA is
\beq \label{rpachemical}
D_c(\vec q,i\Omega)=\frac{1}{k_0\frac{|\Omega|}{|\vec q|}+\chi_dq^2+\rho_s}
\eeq
where $k_0$ is of the order of a typical Fermi momentum of the spinon Fermi surface, and $\chi_dq^2$ is the diamagnetic term. Upon approaching the quantum critical point from the Fermi liquid side, the superfluid density vanishes as $\rho_s\sim\xi^{-z}\sim(g-g_c)^{z\nu}$ up to a logarithmic correction that we will ignore \cite{Sachdev2007}, with $z=2$ and $\nu=\frac{1}{2}$ in the universality class of a dilute bose gas. After the system passes the critical point and enters the spin liquid Mott insulator phase, the RPA gauge field propagator in Coulomb gauge becomes
\beq
D_c(\vec q,i\Omega)=\frac{1}{k_0\frac{|\Omega|}{|\vec q|}+\left(\chi_d+\frac{1}{\Delta}\right)q^2}
\eeq
where $\Delta$ is of the order of the boson gap.

Now let us examine the effects of interlayer interactions. Similar to the warm-up example in Sec. \ref{secom}, the interlayer interactions should also obey the gauge invariance within each layer. In the present case the gauge field structure is $U(1)$.  Apart from electron tunneling between layers, the other most important interlayer interactions here include the coupling between energy densities of different layers of the form (\ref{interlayercoupling1}) and the coupling between the energy density of one layer and the collective excitation of the spinon Fermi surface of the other layer of the form (\ref{interlayercoupling2}). Since the chemical potential tuned Mott transition  is in the universality class of a dilute bose gas, the interaction (\ref{interlayercoupling1}) is already seen to be irrelevant due to similar reasons as discussed in Sec. \ref{secom}. Also, although the spinons are in their own non-Fermi liquid state at the critical point, the factor $\Pi(\vec q,\omega)$ obtained by integrating the spinons out still has the form given by (\ref{densitycorrelator})\cite{KimFurusakiWenEtAl1994}. Again, because the dynamical exponent $z=2$, this interaction is also irrelevant.

However, due to the presence of a gapless $U(1)$ gauge field, there is a coupling of the form
\beq \label{interlayercoupling3}
\delta\mc{L}_3=\sum_{\alpha\beta}\int d\tau d^2x g_{3\alpha\beta}\left(\nabla\times\vec a_\alpha\right)\cdot\left(\nabla\times\vec a_\beta\right)
\eeq
In momentum space, this coupling has the form $\delta\mathcal{L}_3=\sum_{\alpha\beta}\int d\omega d^2q g_{3\alpha\beta}q^2\vec a(\vec q,i\omega)\cdot\vec a(-\vec q,-i\omega)$. At the critical point, $\rho_s=0$, and comparing (18) and the momentum-space representation of $\delta\mathcal{L}_3$, we see this coupling is marginal. As for any physics that only involves quantities within the same layer, the effect of this coupling is merely to modify the effective diamagnetic susceptibility. In particular, under the assumption that the interlayer interactions are weak, this coupling is not able to change the sign of $\chi_d$, so it will not modify the physics qualitatively. Its most important effect on physics involving different layers may be that it can potentially enhance the interlayer pairing (see Appendix \ref{interlayer pairing}). Since in this paper we consistently assume we are in a regime away from any pairing instability, we will also ignore it here and (\ref{interlayercoupling3}) will not modify the physics.

One should in principle also consider the couplings between electric currents, spin currents and spin densities between different layers. The most relevant coupling between electric currents contains two more derivatives compared to (\ref{interlayercoupling1}), so it is even more irrelevant than the latter. The most relevant couplings between spin currents and spin densities are all of the form of four-fermion interaction, which is not relevant for similar reasons as in the case of the orthogonal metal transition.

Of course potentially the most important interlayer coupling is through electron tunneling between the different layers.  We  can discuss it within the framework introduced to analyze the same issue for the orthogonal metal transition. As the exponent $\alpha < 0$ here as well the interlayer hopping will scale to zero at low energies.

So we see that at the quantum critical point of a chemical potential tuned continuous Mott transition, no interlayer interaction is relevant with respect to the decoupled fixed point. Going into the spin liquid Mott insulator side, because the bosons are gapped, the interlayer interactions that involve charges become more irrelevant while other interactions stay as irrelevant as they are at the critical point, so different layers are still decoupled. Therefore, we conclude that at the quantum critical point and in the Mott insulator phase of a chemical potential tuned Mott transition, the system behaves as a stack of many decoupled 2D layers and hence exhibits dimensional decoupling. Again, we notice the emergent gauge invariance plays an important role in obtaining this phenomenon.

\subsection{Bandwidth controlled continuous Mott transition}

If the electron filling is fixed to be at half, we can access the bandwidth controlled Mott transition by tuning the ratio of the electron bandwidth to the interaction strength, which can be done, for example, by tuning the pressure. The effective field theory of this phase transition is similar to (\ref{eftc}), except that now the bosons are relativistic and described by the following Lagrangian
\beq
\mc{L}_b=|(\partial_\mu-ia_\mu)b|^2+V(|b|^2)
\eeq
because of the emergent particle-hole symmetry of the boson. This difference has been realized since the study of the transition between a bosonic Mott insulator and a superfluid \cite{FisherWeichmanGrinsteinEtAl1989}.

\begin{figure}[h!]
  \centering
  \includegraphics[width=0.3\textwidth]{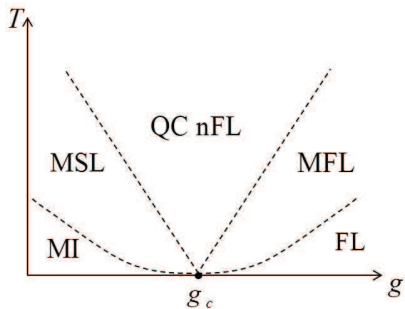}\\
  \caption{Phase diagram and crossover structure of the bandwidth controlled Mott transition. The $g<g_c$ side corresponds to a spin liquid Mott insulator (MI), and the $g>g_c$ side corresponds to a Fermi liquid (FL) metal. The quantum critical (QC) regime is highly non-Fermi liquid like. In crossing over to the Fermi liquid, the system has to go through an intermediate marginal Fermi liquid regime (MFL). In crossing over to the Mott insulator, the system has to go through a marginal spinon liquid (MSL) regime, but this regime will not be discussed in this paper.}\label{CMT-P diagram}
\end{figure}

As in the case of the chemical potential tuned Mott transition, the boson sector is again dynamically decoupled from the spinon-gauge-field sector, and the transition is therefore of the universality class of the 3D XY model. Using this, it is shown that at the quantum critical point the system is non-Fermi liquid like and electron spectral function is
\beq
\mc{A}_c(\vec k,\omega)\sim\frac{\omega^\eta}{\ln\frac{\Lambda}{\omega}}f\left(\frac{\omega\ln\frac{\Lambda}{\omega}} {v_Fk_\parallel}\right)
\eeq
where $\eta$ is the anomalous dimension of 3D XY model and the universal function $f$ is
\beq
f(x)=\left(1-\frac{1}{x}\right)^\eta\theta(x-1)
\eeq
Fitting this spectral function into the form (\ref{cfsscaling}), we find $\alpha=-\eta$ and $z=1^+$, with the understanding that expression such as $\omega^{\frac{1}{z}}$ should be interpreted as $\omega\ln\frac{1}{\omega}$. Again, we see (\ref{zalphabound}) is satisfied.

In addition, the crossover out of QC regime to the Fermi liquid regime again involves an intermediate regime. Choosing the boson phase stiffness $\rho_s$ as the characteristic energy scale of the Fermi liquid, when the temperature is such that $T\gg\rho_s$, the system is in its QC regime and displays non-Fermi liquid behaviors. When the temperature is decreased so that $T\ll\rho_s$, the bosons behave as if they already condensed ({\em i.e} Higgsed) , but the spinon-gauge-field sector does not feel the Higgs effect until the temperature is further lowered to the order of $\rho_s^2$. In this intermediate regime, the system behaves as a marginal Fermi liquid (MFL) state that was originally proposed by Varma {\it et al.} to describe the optimally doped cuprates \cite{VarmaLittlewoodSchmitt-RinkEtAl1989}. Only when the temperature is further lowered so that $T\ll\rho_s^2$ does the spinon-gauge system
notice the the boson condensation  and the system behaves as a Fermi liquid. The difference in the powers of $\rho_s$ below which scale a Fermi liquid results in the chemical potential tuned and bandwidth tuned Mott transitions reflects that the two transitions are in different universality classes.

In this case on the Fermi liquid side the RPA gauge field propagator in Coulomb gauge is
\beq \label{rpabandwidth}
D_b(\vec q,i\Omega)=\frac{1}{k_0\frac{|\Omega|}{|\vec q|}+\sigma_0\sqrt{\Omega^2+\vec q^2}P\left(\frac{\sqrt{\Omega^2+\vec q^2}}{\rho_s}\right)}
\eeq
where the universal function $P(x)$ satisfies $\lim_{x\rightarrow 0}P(x)\sim\frac{1}{x}$ and $\lim_{x\rightarrow\infty}P(x)=1$, and $\sigma_0\sim\frac{e^2}{h}$ is a universal conductance. On the Mott insulator side, it becomes
\beq \label{rpabandwidthmott}
D_b(\vec q,i\Omega)=\frac{1}{k_0\frac{|\Omega|}{|\vec q|}+\sigma_0\sqrt{\Omega^2+\vec q^2}Q\left(\frac{\sqrt{\Omega^2+q^2}}{\Delta}\right)}
\eeq
where $\Delta$ is on the order of the boson gap and the universal function $Q(x)$ satisfies $\lim_{x\rightarrow 0}Q(x)\sim x$ and $\lim_{x\rightarrow \infty}Q(x)=1$.

Suppose we have a stack of such 2D layers, let us now examine the effects of interlayer interactions on the decoupled fixed point. These interlayer interactions must be invariant under the local $U(1)$ gauge transformation within each layer.  We will follow closely the strategy used in previous sections. Interlayer electron tunneling is potentially the most important coupling. Because in this case $\alpha=-\eta<0$, according to the similar arguments in the previous sections it is irrelevant. The other important interlayer interactions are again the coupling between energy densities on different layers and the coupling between the energy density in one layer and the collective excitations of the spinon Fermi surface in another layer, which still has the form of (\ref{interlayercoupling1}) and (\ref{interlayercoupling2}), respectively. However, because this transition is in another universality class, the previous argument should be modified. In particular, because the energy density $|b|^2$ has scaling dimension $3-\frac{1}{\nu}$, the scaling dimension of $g_{1\alpha\beta}$ is $\frac{2}{\nu}-3$. It is known that $\nu>\frac{2}{3}$ for the 3D XY model, so $g_{1\alpha\beta}$ has negative scaling dimension and (\ref{interlayercoupling1}) is irrelevant. On the other hand, integrating out the degrees of freedom from the spinons in (\ref{interlayercoupling2}), the most relevant interaction we get is again of the form (\ref{fermionsintegratedout}). Since we have dynamical exponent $z=1$ in this case, the scaling dimension of coupling constant of this resulting interaction is the same as that of the coupling between energy densities, which is negative as discussed above. Therefore, this interaction is also irrelevant.

Notice in this case, the coupling of the form (\ref{interlayercoupling3}) is simply irrelevant at the critical point, because there the gauge field propagator (\ref{rpabandwidth}) is
\beq \label{rpabandwidthcritical}
D_b(\vec q,i\Omega)=\frac{1}{k_0\frac{|\Omega|}{|\vec q|}+\sigma_0\sqrt{\Omega^2+\vec q^2}}
\eeq
For the gauge field the most important fluctuations are the modes with $\Omega\sim q^2\ll q$, so we can further approximate (\ref{rpabandwidthcritical}) as
\beq
D_b(\vec q,i\Omega)=\frac{1}{k_0\frac{|\Omega|}{|\vec q|}+\sigma_0|\vec q|}
\eeq
The coupling of the form (\ref{interlayercoupling3}) contains two spatial derivatives, so it is irrelevant. However, because of the structure of (\ref{rpabandwidthmott}), as in the case of the chemical potential tuned Mott transition, (\ref{interlayercoupling3}) is marginal deep in the Mott insulator side. But similar arguments as there indicates this coupling does not modify the physics.

For similar reasons as in the case of chemical potential tuned Mott transition, the couplings between charge currents, spin currents and spin densities in different layers are also not relevant. As for interlayer electron tunneling, the critical Fermi surface in this problem has $\alpha = -\eta  < 0$. Thus the interlayer hopping will renormalize to zero at the critical point at low energies. In the Mott insulator phase, because the boson is gapped, the interlayer interactions involving charge become more irrelevant while other interactions stay as irrelevant as they are at the critical point, so different layers are still decoupled.

Therefore, we conclude that  none of the interlayer interactions at the quantum critical point of the bandwidth controlled Mott transition and in the Mott insulator phase is relevant, and the system exhibits dimensional decoupling. It is again the emergent gauge invariance that constrains the possible forms of the interlayer interactions and give rise to this phenomenon.

\section{Interlayer conductivity} \label{secconductivity}

In this section we  derive the interlayer electric conductivity $\sigma$  in perturbation theory in the electron tunneling. As is well known, this leads to a formula for  the interlayer conductivity  in terms of the electron spectral function in each layer. We then we apply the formula to the QC regimes of the various phase transitions discussed above, and show in all three cases $\sigma\rightarrow 0$ as $T\rightarrow 0$. This is of course consistent with that the interlayer tunneling is irrelevant at the decoupled fixed point.

 Suppose the total Hamiltonian of the layered three dimensional system is
\beq \label{multilayerHamiltonian}
H=\sum_aH_a+H_{\rm tunneling}
\eeq
where $H_a$ is the Hamiltonian for the $a$th layer that may take the form of (\ref{Hubbard}), and the tunneling Hamiltonian $H_{\rm tunneling}$ takes the following specific form
\beq
H_{\rm tunneling}=-t\sum_{a,i}\left(c_{i,a+1}^\dag c_{i,a}+{\rm h.c.}\right)
\eeq
where $a$ labels the layer and $i$ labels the position of the site on a given layer.

As shown in Appendix \ref{appkubo}, the interlayer DC electric conductivity to the second order of interlayer tunneling amplitude $t$ is
\beq \label{masterformula}
\sigma=-\pi N(ted)^2\sum_k\int d\omega\left(\mc{A}(\vec k,\omega)\right)^2\frac{df}{d\omega}
\eeq
where $N$ is the number of layers, $d$ is the interlayer spacing, $\mc{A}(\vec k,\omega)$ is the electron spectral function on each layer, and $f(x)=\frac{1}{e^{\beta x}+1}$ is the Fermi-Dirac function.

For systems with a critical Fermi surface, in the quantum critical regime where the zero temperature correlation length $\xi$ is so large that it drops out from the universal function, we can write the scaling form of the spectral function (\ref{cfsscaling}) as
\beq \label{scalingspectralqc}
\mc{A}(\vec k,\omega)=\frac{c_0}{|\omega|^{\frac{\alpha}{z}}}F_c\left(\frac{\omega}{T},\frac{c_1k_\parallel}{T^{\frac{1}{z}}}\right)
\eeq
with another universal function $F_c$ simply related to the original universal function $F$ in (\ref{cfsscaling}) via $F_c(x,y)=F(x,y,\infty)$.
Applying (\ref{scalingspectralqc})to (\ref{masterformula}), we get
\beq \label{sigmascalingqc0}
\sigma\sim\int d\theta\frac{c_0(\theta)^2}{c_1(\theta)}T^{\frac{1-2\alpha(\theta)}{z(\theta)}}
\eeq
where $\theta$ denotes the angular position of a patch on the Fermi surface, and the integral is over all these patches. Therefore, as long as $\alpha<\frac{1}{2}$ for each patch, the interlayer electric conductivity vanishes as the temperature goes to zero. Notice the condition for the interlayer electron hopping to be irrelevant, i.e. $\alpha<0$, is stronger than this condition, which is of course expected. Both conditions are satisfied for all three cases discussed above, and this confirms our previous assertion that $\sigma\rightarrow 0$ as $T\rightarrow 0$.

Recall that we have a general inequality (\ref{zalphabound}) for a system with a critical Fermi surface, from which we obtain
\beq
\frac{1-2\alpha}{z}\geqslant\frac{1}{z}-2>-2
\eeq
This implies for such systems, in this perturbative regime, the scaling behavior of the interlayer electric conductivity with respect to temperature cannot be more singular than that of a Fermi liquid
\beq
\sigma\lesssim\frac{1}{T^2}
\eeq

Now we apply (\ref{sigmascalingqc0}) to the examples discussed in the previous sections. For all examples, the exponents $\alpha$ and $z$ do not depend on its angular position $\theta$ (even though the non-universal constants $c_0$ and $c_1$ can depend on $\theta$ in general), and (\ref{sigmascalingqc0}) simplifies to {\footnote {One can similarly get the zero temperature AC conductivity to be $\sigma(\omega)\sim\omega^{\frac{1-2\alpha}{z}}$.}}
\beq \label{sigmascalingqc}
\sigma\sim T^{\frac{1-2\alpha}{z}}
\eeq

For the QC regimes of  the chemical potential tuned Mott transition (and the related problem of the orthogonal metal transition) , we have $\alpha=-1$ and $z=2$. Therefore
\beq
\sigma\sim T^{\frac{3}{2}}
\eeq
For the QC regime of the bandwidth controlled Mott transition, we have $\alpha=-\eta$ and $z=1^+$, so up to logarithms we have
\beq \label{sigmascalingbandwidthcritical}
\sigma\sim T^{1+2\eta}
\eeq

Before continuing, we comment on the in-plane electric conductivity in these QC regimes. For orthogonal metals, as discussed before, because the fermion $f$ carries charge and it is in its Fermi liquid state, the in-plane electric conductivity is Fermi liquid like. In the QC regime of the chemical potential tuned Mott transition, the behavior of the in-plane conductivity is non-Fermi liquid like. In particular, this conductivity behaves as $\ln\left(\frac{1}{T}\right)$ at the lowest temperatures according to the conventional slave-particle theories \cite{SenthilVojtaSachdev2004}. As for the QC regime in the bandwidth controlled Mott transition, if $g>g_c$, the in-plane {\it resistivity} has a universal crossover from a finite value to the impurity-induced residual resistivity of the Fermi liquid \cite{Senthil2008a,Witczak-KrempaGhaemiSenthilEtAl2012}.

\section{Crossovers out of criticalities} \label{seccrossover}

As reviewed in Sec. \ref{seccmt}, a very interesting feature of the continuous Mott transitions is that the crossovers out from the QC regimes to the Fermi liquid regime involve an intermediate regime: there is an IFL regime for chemical potential tuned Mott transition and an MFL regime for bandwidth controlled Mott transition.  It is clearly interesting to  study the  interlayer electric transport properties in these regimes.

As long as at the lattice level the interlayer interactions are much weaker than the intralayer interactions, it may still be legitimate to apply (\ref{masterformula}) to calculate the interlayer electric conductivity. However, unlike that in the quantum critical regime where the zero temperature correlation length $\xi$ drops out from the universal function and we can get the scaling behavior of the conductivity simply by using the scaling form of the electron spectral function $\mc{A}(\vec k,\omega)$, here we no longer have a simple scaling form for it and we need to calculate it explicitly.

This is done by recalling that the physical electron operator is written as
\beq
c_{i\sigma}=f_{i\sigma}b_i
\eeq
In both IFL and MFL regimes, despite the Mermin-Wagner theorem (stateing that there is no true long-range order at any finite temperature in 2D), as long as $T\ll\rho_s$, the correlation length and correlation time are still extremely large, and the bosons behave as if they already condense. Therefore, to calculate the electron spectral function, we only need to calculate the spinon spectral function and multiply it by the condensate magnitude $|\la b\ra|^2\sim(g-g_c)^{2\beta}$, where $\beta$ is the order parameter exponent of the transition.

The spinon spectral function is determined by the imaginary part of the spinon Green's function
\beq
\mc{A}_f(\vec k,\omega)=-\frac{1}{\pi}{\rm Im}\mc{G}_f(\vec k,i\omega\rightarrow\omega+i\eta)
\eeq
Writing the (real frequency) spinon Green's function in terms of spinon self-energy $\Sigma_f(\vec k,\omega)$, we get
\beq
\mc{G}_f(\vec k,\omega)=\frac{1}{i\omega-\epsilon_f(\vec k)-\Sigma_f(\vec k,\omega)}
\eeq
with the bare spinon dispersion
\beq \label{dispersion}
\epsilon_f(\vec k)=v_Fk_\parallel+\kappa k_\perp^2
\eeq
where $v_F$ is a (finite) bare Fermi velocity, and $\kappa$ is determined by the Fermi surface curvature. So the spinon spectral function is given by
\beq \label{spinonspectral}
\begin{split}
&\mc{A}_f(\vec k,\omega)=\\
&\quad
-\frac{1}{\pi}\frac{\Sigma_f''(\vec k,\omega)}{\left(\omega-\epsilon_f(\vec k)-\Sigma_f'(\vec k,\omega)\right)^2+\left(\Sigma_f''(\vec k,\omega)\right)^2}
\end{split}
\eeq
where $\Sigma_f'(\vec k,\omega)$ and $\Sigma_f''(\vec k,\omega)$ are the real and imaginary parts of $\Sigma_f(\vec k,\omega)$, respectively.

\subsection{IFL in a chemical potential tuned Mott transition}

\begin{figure}[h!]
  \centering
  \includegraphics[width=0.2\textwidth]{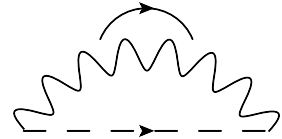}\\
  \caption{Spinon self-energy. The wavy line represents a gauge boson with momentum and frequency $(\vec q,i\Omega)$, and the dashed line represents a spinon with momentum and frequency $(\vec k-\vec q,i(\omega-\Omega))$.}\label{spinon self-energy diagram}
\end{figure}

According to (\ref{spinonspectral}), to calculate the spinon spectral function, we need to calculate the spinon self-energy. In the IFL regime, the imaginary frequency self-energy of the spinons near a certain patch on the spinon Fermi surface is given by (see Fig. \ref{spinon self-energy diagram})
\beq \label{spinonselfenergy}
\begin{split}
&\Sigma_f(\vec k,i\omega)\\
&\qquad
=v_{F}^2T\sum_{\Omega_n}\int_{\vec q}D_c(\vec q,i\Omega)\mc{G}_f^{(0)}(\vec k-\vec q,i(\omega-\Omega))
\end{split}
\eeq
where the gauge field propagator $D_c(\vec q,i\Omega_n)$ takes the form of (\ref{rpachemical}), and the bare spinon propagator $\mc{G}_f^{(0)}(\vec k,i\omega)$ is given by
\beq
\mc{G}_f^{(0)}(\vec k,i\omega)=\frac{1}{i\omega-\epsilon_f(\vec k)}
\eeq
The summation is over the bosonic Matsubara frequencies $\Omega=2\pi nT$, $n=0,\pm 1, \pm 2, \cdots$.

Due to the structure of the bare spinon propagator, the important region of integral involves $q_\parallel\sim q_\perp^2\ll q_\perp$, so we can ignore the $q_\parallel$ dependence in the gauge field propagator $D_c(\vec q,i\Omega)$. After this, the integral over $q_\parallel$ can be done and we find the spinon self-energy does not have a singular dependence on $k_\parallel$. This implies, because of the form of the spinon spectral function (\ref{spinonspectral}) and the form of the interlayer electric conductivity (\ref{masterformula}), having $\Sigma_f''(\vec k,\omega)$ is sufficient to obtain $\sigma$.

We calculate $\Sigma_f''(\vec k,\omega)$ and find its leading singular part is (see Appendix \ref{appcalculations})
\beq \label{spinonselfenergyc}
\Sigma_f''(\vec k,\omega)\sim
\left\{
\begin{array}{lr}
-\frac{T}{\sqrt{\rho_s}}, &|\omega|\lesssim T\\
-|\omega|^{\frac{2}{3}}, &|\omega|\gtrsim T
\end{array}
\right.
\eeq
Plugging this result into (\ref{spinonspectral}) and (\ref{masterformula}), we find the leading dependence of $\sigma$ on the temperature in IFL regime to be
\beq \label{sigmaifl}
\sigma\sim t^2|\la b\ra|^4T^{-\frac{2}{3}}
\eeq

\begin{figure}[h!]
  \centering
  \includegraphics[width=0.23\textwidth]{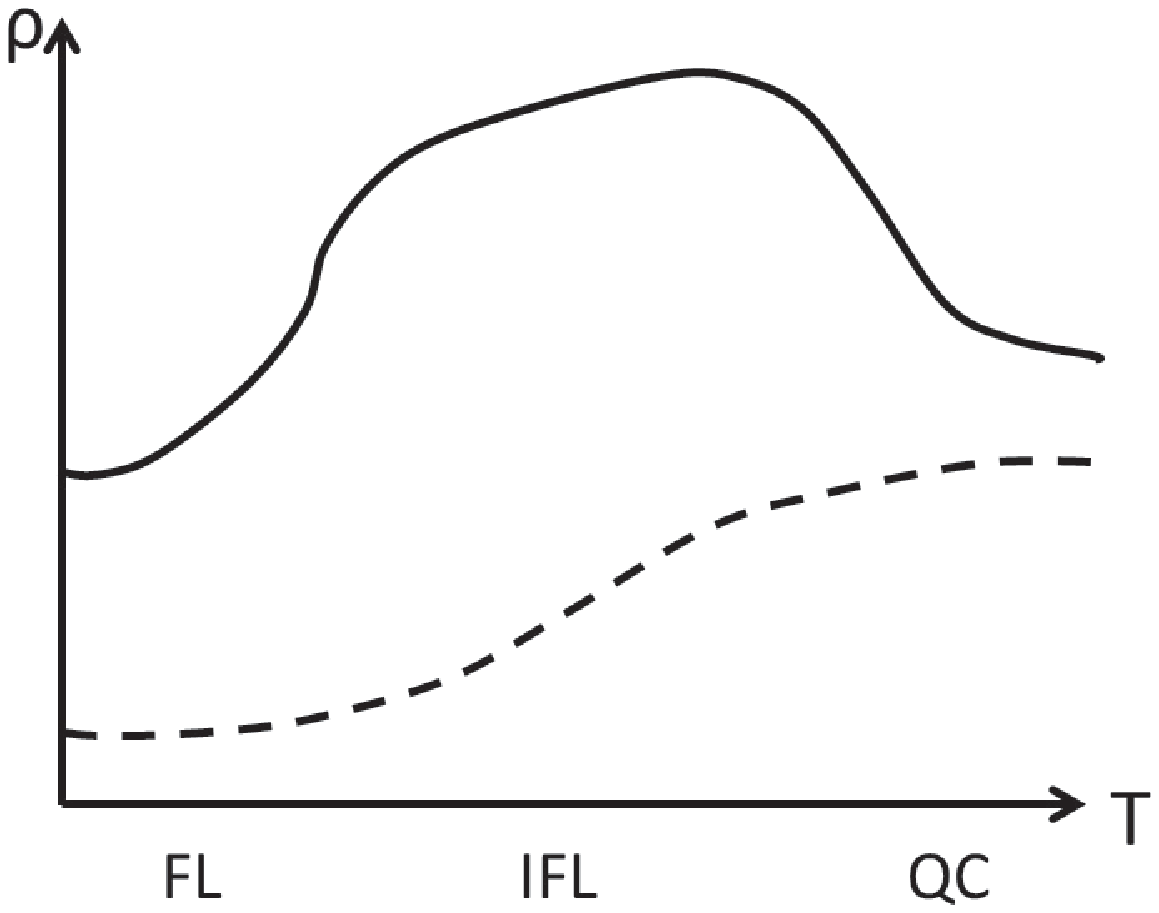}
  \includegraphics[width=0.23\textwidth]{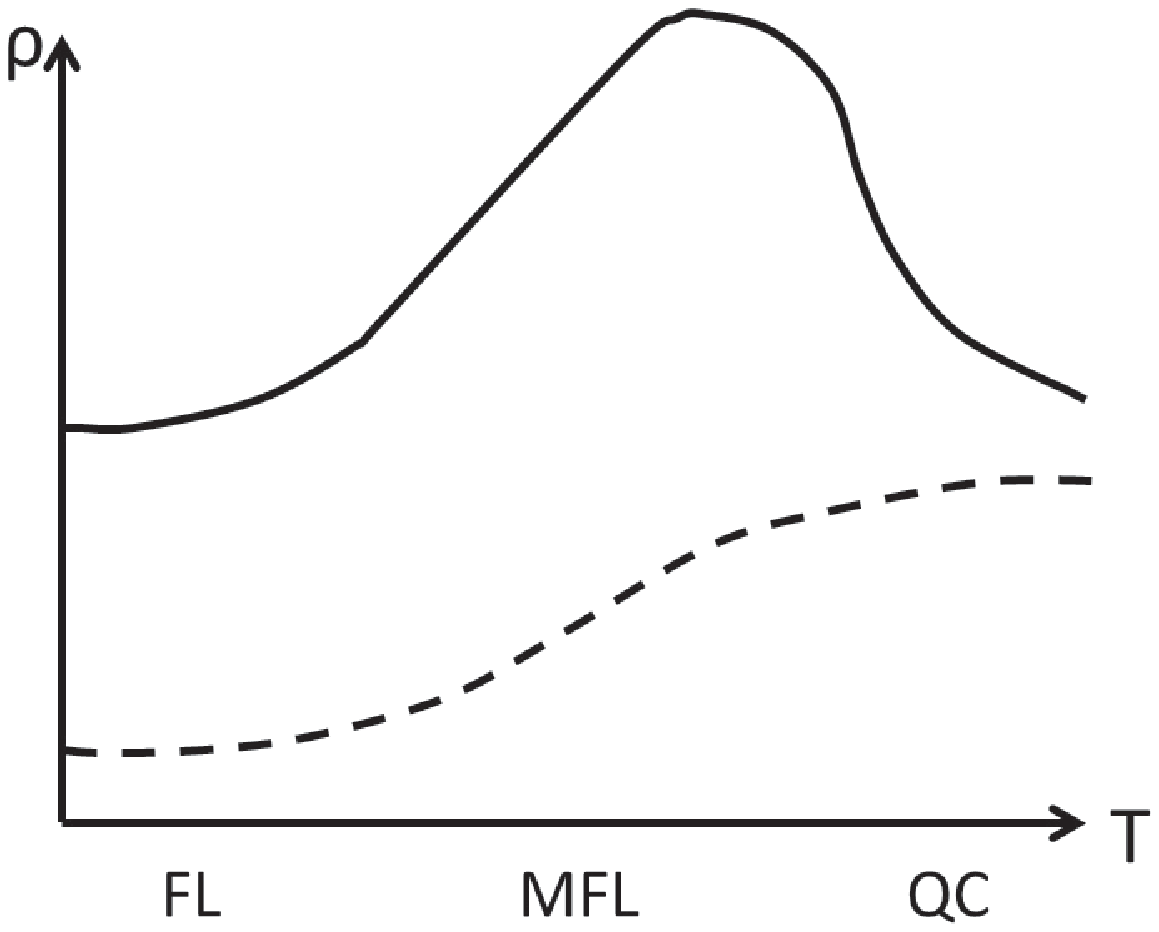}\\
  \caption{Schematic crossover behaviors of the resistivity with respect to temperature for chemical potential tuned Mott transition (left) and for bandwidth controlled Mott transition (right). The solid line represents the out-of-plane conductivity and the dashed line represents the in-plane conductivity.}\label{fig:crossover}
\end{figure}

This result means in the IFL regime, if the temperature is fixed, when we get closer and closer to the QC regime, the interlayer electric conductivity decreases as $(g-g_c)^{4\beta}$. On the other hand, if we fix $g$, which experimentally corresponds to fixing the doping, as the temperature is lowered, the interlayer electric conductivity increases as $T^{-\frac{2}{3}}$, so the out-of-plane transport appears to be conducting. As discussed in Ref. \onlinecite{SenthilLee2009}, the in-plane conductivity of the IFL regime is non-Fermi liquid like, which behaves as $T^{-\frac{4}{3}}$. Combining these and previous results for the QC regime, this implies on the $g>g_c$ side, in the QC regime, the in-plane conductivity is non-Fermi liquid like, while the out-of-plane electric transport appears to be insulating. When the temperature is decreased and the system enters the IFL regime, the in-plane conductivity is still non-Fermi liquid like, but the out-of-plane electric transport already seems to be conducting but non-Fermi liquid like. If the temperature is further decreased, both the in-plane and out-of-plane electric transports behave like a Fermi liquid (see Fig. \ref{fig:crossover}). This phenomenon can be viewed as an experimental signature of the crossover structure displayed in Fig. \ref{CMT-mu diagram}.

\subsection{MFL in a bandwidth controlled Mott transition}

Similar to the chemical potential tuned Mott transition, we again just need the imaginary part of the spinon self-energy to calculate $\sigma$. The difference between these two cases is that now the gauge field propagator is given by (\ref{rpabandwidth}). Plugging (\ref{rpabandwidth}) into (\ref{spinonselfenergy}), we find the leading singular part of $\Sigma_f''(\vec k,\omega)$
\beq \label{spinonselfenergyb}
\Sigma_f''(\vec k,\omega)\sim
\left\{
\begin{array}{lr}
-T\ln\frac{T}{\rho_s^2}, &|\omega|\lesssim T\\
-\left(|\omega|+\lambda T\ln\frac{|\omega|}{\rho_s^2}\right), &|\omega|\gtrsim T
\end{array}
\right.
\eeq
with $\lambda\sim\mc{O}(1)$ a non-universal constant. To the best of our knowledge, the result for $|\omega|\lesssim T$ is new. Combining this result, (\ref{spinonspectral}) and (\ref{masterformula}), we get the leading dependence of $\sigma$ on the temperature in MFL regime to be
\beq \label{sigmascalingmfl}
\sigma\sim t^2|\la b\ra|^4\left(T\ln\left(\frac{T}{\rho_s^2}\right)\right)^{-1}
\eeq

If we ignore the logarithmic correction in the above result, just as the case of the IFL regime in the chemical potential tuned Mott transition, when the system gets closer to the QC regime from the MFL regime in the bandwidth controlled Mott transition with temperature fixed, the interlayer electric conductivity becomes smaller and smaller as if it vanishes as $(g-g_c)^{4\beta}$. And if we decrease the temperature with $g$ fixed, which experimentally corresponds to fixing the pressure, the interlayer electric conductivity increases as $T^{-1}$, so the out-of-plane transport appears to be conducting. As shown in Appendix \ref{appcalculations}, the in-plane electric conductivity in the MFL regime behaves as $T^{-2}$, which is Fermi liquid like. Combining these results and the discussion on the QC regime, we see on the $g>g_c$ side, in the QC regime the in-plane resistivity has a universal jump, while the out-of-plane electric transport seems insulating. When the temperature is lowered and the system enters the MFL regime, the in-plane conductivity already behaves as that of a Fermi liquid, while the out-of-plane conductivity is conducting but non-Fermi liquid like. If the temperature is further lowered so that the system is in the Fermi liquid regime, the temperature dependence of the in-plane conductivity does not change qualitatively, and the out-of-plane conductivity eventually becomes Fermi liquid like (see Fig. \ref{fig:crossover}). This phenomenon serves as an experimental signature of the crossover structure displayed in Fig. \ref{CMT-P diagram}.

We end this section by commenting on the validity of the perturbative calculation of $\sigma$. We first discuss the case of the IFL regime. Notice (\ref{masterformula}) is derived perturbatively up to the second order of the interlayer tunneling amplitude $t$, and (\ref{sigmaifl}) shows $\sigma\rightarrow\infty$ as $T\rightarrow 0$, so one may wonder whether the higher order terms in $t$ should be included. However, in the IFL regime, we are working in a temperature regime $T\gg\rho_s^{\frac{3}{2}}$ and we will not go into an arbitrarily low temperature. As argued in Appendix \ref{appvalidity}, in this regime the higher order contributions are expected to be indeed small compared to this leading order contribution, so the perturbative result (\ref{sigmaifl}) is valid.

Now we turn to the MFL regime. Again, in the MFL regime $T\gg\rho_s^2$ and we will not go to an arbitrarily low temperature. As argued in Appendix \ref{appvalidity}, in most parameter regimes of experimental interests, the perturbative calculation is expected to be valid as long as the interlayer electron tunneling amplitude $t$ is small. If the system is extremely close to the quantum critical point, there may be a narrow window where the perturbative calculation breaks down, and the temperature dependence of $\sigma$ is expected to be more singular than (\ref{sigmascalingmfl}), but it should not be more singular than the Fermi liquid form $T^{-2}$. Since this window is very narrow, it may not be too significant experimentally.

\section{Discussion} \label{secdiscussion}

In this paper, we have demonstrated the phenomenon of dimensional decoupling at continuous Mott transitions between a Fermi liquid metal and a Mott insulator in a multilayered quasi-2D system. At low energies, in the Mott insulating phase as well as right at the quantum critical point, the system behaves as a stack of many decoupled 2D layers, while it behaves as a 3D Fermi liquid in the metallic side. Experimentally, for example, this implies the interlayer electric transport will become insulating in the quantum critical regimes of these transitions. We emphasize the importance role that electron fractionalization plays in obtaining dimensional decoupling, and we also point out, under reasonable assumptions, this phenomenon cannot occur in a non-Fermi liquid obtained near a conventional quantum critical point that is associated with a spontaneous breaking of internal symmetries.

By calculating the temperature dependence of the interlayer electric conductivity $\sigma$ induced by electron tunneling between different layers, we have systematically explored the crossover behavior of the interlayer transport in Sec. \ref{seccrossover}. We find for these continuous Mott transitions the interlayer conductivity vanishes as the temperature goes to zero in the QC regimes, which is consistent with that the interlayer electron tunneling is irrelevant at low energies. Intuitively, this is because an electron will be split into a boson and a fermion across the transitions. To have an electron tunnel from one layer to the other, both the boson and the fermion have to tunnel collectively. However, since the boson is to become gapped across the transitions, this process is suppressed. One of the interesting features of these continuous Mott transitions is the existence of an intermediate regime when the system crosses over from the QC regime to the Fermi liquid, and we also derived the scaling behavior of $\sigma$ in these regimes and showed $\sigma$ increases as temperature decreases. In particular, in the IFL regime of the chemical potential tuned Mott transition we find $\sigma\sim T^{-\frac{2}{3}}$, and in the MFL regime of the bandwidth controlled Mott transition we find $\sigma\sim\left(T\ln\frac{1}{T}\right)^{-1}$. This metallic behavior of interlayer transport is because of the partial recombination of the boson and the fermion. When the temperature is low enough so that the system enters the Fermi liquid regime, the boson and the fermion will fully combine and become the original electron, so the system behaves as a coherent 3D metal where both the intralayer and interlayer conductivities behave as $T^{-2}$. Therefore, there will be a coherence peak in the temperature dependence of the interlayer resistivity (see Fig. \ref{fig:crossover}). The position of the peak is around where the system crosses out from the QC regimes into the intermediate regimes, so the deeper the system is in the Fermi liquid side, the higher the temperature of this peak.   These predictions serve as useful guidance to compare this theory to experiments.

Recently out-of-plane transport of a layered quasi-2D doped organic compound has been studied experimentally\cite{OikeSuzukiTaniguchiEtAl2016}. This material is suggested to be a candidate of a doped spin liquid, in the sense that charge and spin will be separated in the absence of doping. As reviewed in Sec. \ref{secintro}, this material shows non-Fermi liquid like transport behavior at ambient pressure, and becomes Fermi liquid like under high enough pressure. So pressure can be viewed to drive this crossover from non-Fermi liquid to Fermi liquid, and the higher the pressure is, the deeper the system is in the Fermi liquid side. It is found that there is indeed a regime where the in-plane transport is non-Fermi liquid like, and the out-of-plane transport behaves insulating at high temperatures but becomes metallic at low temperatures. The coherence peak of the interlayer resistivity seems to occur at a temperature that is small compared to the relevant lattice energy scales, and it shifts to higher temperatures as the pressure is increased. All these features agree with the predictions of our theory on the metallic side qualitatively. Ref. \onlinecite{OikeSuzukiTaniguchiEtAl2016} also suggested, similar to the considerations in this paper, that the metallic behavior of out-of-plane transport when the in-plane transport is non-Fermi liquid like is due to the recombination of charge and spin.

Finally, we note, as discussed in Sec. \ref{secconductivity}, for a layered quasi-2D non-Fermi liquid, as long as the electron spectral function is singular enough ($\alpha>\frac{1}{2}$), the out-of-plane transport may be metallic. This scenario can occur in non-Fermi liquids obtained near a conventional quantum critical point where the transition is driven by fluctuating local order parameter, and a possible example is discussed in Appendix \ref{appisingnematics}. However, in this scenario there is not expected to be a peak in the temperature dependence of the interlayer resistivity as long as the system is in the scaling regime of the transition. Therefore, the observation made in Ref. \onlinecite{OikeSuzukiTaniguchiEtAl2016} is unlikely to fall into this scenario.

\section{Acknowledgement} \label{secaknowledgement}

We acknowledge helpful discussions with Debanjan Chowdhury, K. Kanoda, Samuel Lederer, Adam Nahum, Michael Pretko and Chong Wang.  This work was supported by a  US Department of Energy
grant DE-SC0008739. TS was also partially supported by a Simons Investigator award from the Simons Foundation.

\begin{appendix}

\section{An example that does not display dimensional decoupling: non-Fermi liquid near an Ising-nematic critical point} \label{appisingnematics}

In this appendix we provide an example of non-Fermi liquid that does not display dimensional decoupling: the non-Fermi liquid near an Ising-nematic critical point.

Consider an electronic system that has an instability towards the formation of an Ising ferromagnet, which breaks the $Z_2$ symmetry of spin flips. Such a phase transition can viewed as a realization of the ``Ising-nematic" transition \cite{MetlitskiSachdev2010}. Theoretically, the universal physics of the Ising-Nematic transition in 2+1 dimension is believed to be described by focusing on a pair of patches on the Fermi surface that are parallel to each other. The low-energy effective Lagrangian is
\beq
\begin{split}
\mc{L}_{\rm IN}
&=\sum_{s=\pm}\psi^\dag_{s}\left(\eta\frac{\partial}{\partial\tau}-is\frac{\partial}{\partial x}-\frac{\partial^2}{\partial y^2}-\lambda\phi\right)\psi_s\\
&\qquad
+(\partial_y\phi)^2+r\phi^2
\end{split}
\eeq
where $\psi_s$ with $s=\pm$ are the fermion operators on the two parallel patches, and $\phi$ is the Ising-nematic order parameter that flips sign under the $Z_2$ transformation\cite{MrossMcGreevyLiuEtAl2010,MetlitskiSachdev2010}.

It is found in the QC regime of this transition, the system is very non-Fermi liquid like, and the electron spectral function near this critical point also has the form of (\ref{cfsscaling}). As calculated in the framework of a controlled expansion, in this case $\alpha\approx 0.7$ and $z=3$\cite{MrossMcGreevyLiuEtAl2010}.

Now consider we have a stack of many layers of such systems and examine the effects of interlayer interactions on the decoupled fixed point. First consider the interlayer electric conductivity $\sigma$ induced by interlayer electron tunneling. Because $\alpha\approx0.7$, the arguments in the main text indicates it is relevant. Furthermore, according to (\ref{sigmascalingqc0}), the electrons are not incoherent enough and $\sigma$ actually diverges as the temperature goes to zero.

We can also consider the coupling between the order parameters between different layers. This coupling is symmetric under a global $Z_2$ transformation for all layers and is thus an allowed perturbation, and it does not involve interlayer electron tunneling. It has the following form
\beq
\sum_{\alpha\beta}g_{\alpha\beta}\int d\tau dxdy\phi_\alpha(x,y,\tau)\phi_\beta(x,y,\tau)
\eeq
Following the RG analysis in Ref. \onlinecite{MetlitskiSachdev2010}, the scaling dimensions of various quantities of interests are $[y]=-1$, $[x]=-2$, $[\tau]=-2$ and $[\phi]=2$, from which we can deduce that $[g_{\alpha\beta}]=2$. This means this coupling is strongly relevant.

These results means that the decoupled fixed point is unstable to interlayer interactions in this example, so it does not display the phenomenon of dimensional decoupling.

In fact, under reasonable assumptions, it can be argued that for any non-Fermi liquid obtained near a conventional quantum critical point associated with a phase transition that involves internal symmetry breaking, the coupling between the order parameters on different layers is relevant. To see this, consider a single 2D layer and denote the order parameter associated with this transition by $\mc{O}$. It is reasonable to assume that at the critical point the susceptibility of the order parameter diverges, which implies the scaling form,
\beq
\la\mc{O}(\vec k,i\omega)\mc{O}(\vec k,i\omega)\ra\sim\frac{1}{k^{\delta}}\cdot h\left(\frac{\omega}{k^z}\right)
\eeq
has $\delta>0$, where $h$ is a universal function. In terms of $\delta$, the scaling dimension of the order parameter is $[\mc{O}]=\frac{D-\delta}{2}$, where $D$ is the total scaling dimension of the space-time.

Now consider a stack of many such 2D layers, the interlayer coupling of order parameters $\int d\tau dxdy\mc{O}_i\mc{O}_j$ has scaling dimension $-\delta<0$, so it is relevant and prevents dimensional decoupling. Notice to get this result we have assumed the susceptibility of the order parameter diverges at the critical point, but we cannot prove this must hold for all symmetry-breaking transition in a metallic environment, neither can we find any counterexample.

Before ending this appendix, we note this statement does not imply the theoretic works that treat these systems as a single 2D layer are all incorrect. In fact, as long as the interlayer interactions are much weaker than the intralayer interactions at the lattice level, this treatment is valid unless one goes to extremely low energies.

\section{RG equation of perturbation $\delta\mc{L}_1$} \label{apponeloop}

In this appendix we derive the RG equation of the perturbation $\delta\mc{L}_1$, (\ref{oneloopRG}).

\begin{figure}[h!]
  \centering
  \includegraphics[width=0.4\textwidth]{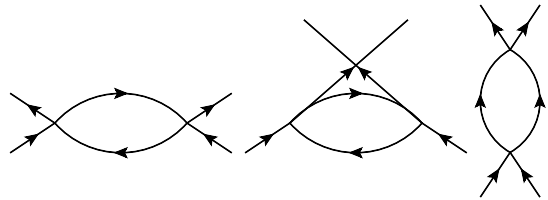}\\
  \caption{Feynman diagrams that contribute to the RG equation of $\delta\mc{L}_1$ to the one-loop order.}\label{oneloopdiagram}
\end{figure}

At tree-level, this perturbation is marginal in two spatial dimensions. Up to one-loop order, there are three distinct Feynman diagrams that contribute to the RG equation of $\delta\mc{L}_1$, as shown in Fig. \ref{oneloopdiagram}. To obtain the RG equation, we will carry out a Wilsonian procedure by integrating out the fast modes with momenta between the old and new cutoff scales, $\Lambda$ and $\Lambda e^{-l}$, respectively, where $e^{-l}$ is a scaling factor. We do not put any cutoff for the frequency. The first two diagrams vanish after the frequency integral is carried out, which, physically speaking, is because for non-relativistic bosons there need to be at least two particles to have any interaction at all and the ground state at the unperturbed critical point is a vacuum with no particle. The third diagram does not vanish, and different $g_{ab}$'s do not mix from this diagram. For the purpose of obtaining the RG equation, we can set the external momenta and frequencies to be zero, then the third diagram gives
\beq
\begin{split}
&-\frac{g_{1\alpha\beta}^2}{(2\pi)^3}\int_{\Lambda e^{-l}}^\Lambda d^2k\int_{-\infty}^\infty d\omega\frac{1}{i\omega-\frac{k^2}{2m_b}}\frac{1}{-i\omega-\frac{k^2}{2m_b}}\\
&\qquad\qquad
=-\mc{C}g_{1\alpha\beta}^2\cdot l
\end{split}
\eeq
with $\mc{C}=\frac{m_b}{2\pi}$. Therefore, the one-loop RG equation of $g_{ab}$ is
\beq
\frac{dg_{1\alpha\beta}}{dl}=-\mc{C}g_{1\alpha\beta}^2
\eeq
In fact, this RG equation is valid up to arbitrary order of perturbation, which is again related to the fact that in this model there need to be at least two particles so that there is interaction at all and the ground state at the critical point is the vacuum with no particle\cite{Sachdev2007,FisherWeichmanGrinsteinEtAl1989}.

\section{Electron spectral function at criticality of the chemical potential tuned Mott transition} \label{appelecspecfnchemical}

In this appendix we calculate the electron spectral function at the critical point of the chemical potential tuned Mott transition.

Because the electron vertex is not singularly enhanced \cite{Senthil2008a}, we can get the electron spectral function by convolving the boson spectral function
\beq \label{bosonchemicalqc}
\mc{A}_b(\vec k,\omega)\sim\delta\left(\omega-\frac{\vec q^2}{2m_b}\right)
\eeq
and the spinon spectral function
\beq
\mc{A}_f(\vec k,\omega)\sim\frac{\omega^{\frac{2}{3}}}{\left(\lambda\sin\frac{\pi}{3}\omega^{\frac{2}{3}}-\epsilon^f_{\vec k}\right)^2+\lambda^2\cos^2\frac{\pi}{3}\omega^{\frac{4}{3}}}
\eeq
where spinon dispersion is taken to be in the form of (\ref{dispersion}).

The electron spectral function is then
\beq \label{convolution}
\mc{A}(\vec k,\omega)=\int_0^\omega d\Omega\int_{\vec q}\mc{A}_b(\vec q,\omega-\Omega)\mc{A}_f(\vec k-\vec q,\Omega)
\eeq

From the expressions of spectral functions of boson and spinon, we see in the low-energy and low-momentum regime the important region of integral is $\omega\sim\Omega\sim q_\parallel^2\sim q_\perp^2\sim k_\parallel^2$. Then in the spinon spectral function, $k_\parallel-q_\parallel$ is much larger than all other terms, and we can approximate it by a delta function
\beq \label{spinonchemicalqc}
\mc{A}_f(\vec k-\vec q,\Omega)
\sim\delta(k_\parallel-q_\parallel)
\eeq
Now plug (\ref{bosonchemicalqc}) and (\ref{spinonchemicalqc}) into (\ref{convolution}), we can do the integral and get
\beq \label{elecspectch}
\begin{split}
\mc{A}(\vec k,\omega)
\sim&|\omega|^{\frac{1}{2}}f\left(\frac{2m_b\omega}{k_\parallel^2}\right)
\end{split}
\eeq
with
\beq
f(x)=\left(1-\frac{1}{x}\right)^{\frac{1}{2}}\theta(x-1)
\eeq
Fitting into the scaling form (\ref{cfsscaling}), we have $\alpha=-1$ and $z=2$. Notice the electron spectral function (\ref{elecspectch}) has the same form as (\ref{elecspectom}), the electron spectral function at the critical point of a $Z_4$ orthogonal metal transition.

\section{Effects on the interlayer pairing of (\ref{interlayercoupling3})} \label{interlayer pairing}

In this appendix we discuss the effects on the interlayer pairing of the coupling (\ref{interlayercoupling3}). First consider a circular Fermi surface in a single layer, then the dominate pairing occurs between opposite patches on the Fermi surface. Because these opposite patches carry opposite electric currents, the Amperean force between them is repulsive, which suppresses pairing \cite{MrossMcGreevyLiuEtAl2010,MetlitskiMrossSachdevEtAl2015}. Now suppose we have many layers coupled by (\ref{interlayercoupling3}), the dominate interlayer pairing is among patches on different layers that have parallel norms. If $g_{3\alpha\beta}$ is positive (negative), the magnetic fields on the $\alpha$-th layer and on the $\beta$-th layer will favor to have opposite (same) signs. When they have opposite (same) signs, the Amperean force between patches on the two layers with parallel norms will be attractive (repulsive). Therefore, (\ref{interlayercoupling3}) will enhance interlayer pairing if $g_{3\alpha\beta}>0$ and will suppress it if $g_{3\alpha\beta}<0$. {\footnote{Similar line of thinking implies that for multiple layers of composite Fermi liquids interacting with repulsive interlayer density-density interactions, the system is unstable against interlayer pairing. This is consistent with the result in Ref. \onlinecite{BonesteelMcDonaldNayak1996}.}}

In the rest of this appendix, we will apply an RG analysis to justify the above physical picture. It is convenient to write (\ref{interlayercoupling3}) in the frequency-momentum space
\beq
\delta\mc{L}_3=\sum_{\alpha\beta}\int_{\vec q,\omega}g_{3\alpha\beta}q^2\cdot a_\alpha(\vec q,i\omega)a_\beta(-\vec q,-i\omega)
\eeq
and introduce the momentum modes along the $z$ direction for the gauge fields:
\beq
a_{q_z}(\vec q,i\omega)=\frac{1}{\sqrt{N}}\sum_\alpha a_\alpha(\vec q,i\omega)e^{-iq_zz_\alpha}
\eeq
Then at the critical point of the chemical potential tuned Mott transition or in the spinon Fermi surface phase, the gauge field propagator can be diagonalized in the $q_z$ basis as
\beq
D_{q_z}(\vec q,i\omega)=\frac{1}{k_0\frac{|\omega|}{|q|}+q^2\left(\chi_d+\mc{E}(q_z)\right)}
\eeq
with the dispersion along the $z$ direction $\mc{E}(q_z)=\sum_{\alpha\beta}g_{3\alpha\beta}e^{iq_z(z_\alpha-z_\beta)}$.

The coupling between the gauge field and the spinons on the $\alpha$-th layer should also be properly written in the $q_z$ basis as
\beq
a_\alpha\bar f_\alpha f_\alpha=\left(\frac{1}{\sqrt{N}}\sum_{q_z}a_{q_z}(\vec q,i\omega)e^{iq_zz_\alpha}\right)\bar f_\alpha f_\alpha
\eeq
As we can see, each $q_z$ mode of the gauge field and the spinons on each layer are all coupled, but there is a factor $e^{iq_zz_\alpha}$ in the coupling constant. Due to the oscillating nature of this factor, different layers may appear to have opposite ``charges" for a given $q_z$ mode, which can potentially induce attractive interlayer four-fermion interaction and enhance the interlayer pairing instability. We note that similar phenomenon has been studied in the context of bilayer composite Fermi liquids \cite{BonesteelMcDonaldNayak1996}.

An RG approach that is suitable to study the effect of pairing in the presence of a gauge field in 2D has been recently developed in Ref. \onlinecite{MetlitskiMrossSachdevEtAl2015}. It is pointed out, the renormalization of the BCS channel scattering amplitude, $V$, comes from two aspects. The first is the conventional contribution arising from renormalizing the thickness of the momentum shell around the Fermi surface being considered to zero, which contributes to the beta function of $V$ by an amount $-V^2$. The second is from the interactions among different patches of the Fermi surface generated by integrating the high momentum modes of the gauge field as the sizes of the patches are renormalized to zero, and this contributes a positive constant to the beta function.

This RG procedure can be straightforwardly generalized into our multilayered system. Denote the interlayer BCS scattering amplitude of the fermions on the $\alpha$-th and the $\beta$-th layers by $V_{\alpha\beta}$, the aforementioned first type of contribution remains to be $-V_{\alpha\beta}^2$, but for the second type of contribution we need to add all $q_z$ modes with each of them multiplied by the oscillating exponential factor $e^{iq_z(z_\alpha-z_\beta)}$. Borrowing the result from Ref. \onlinecite{MetlitskiMrossSachdevEtAl2015}, the beta function of $V$ in the weak interlayer coupling regime can be calculated to be
\beq
\begin{split}
\frac{dV_{\alpha\beta}}{dl}
=-\epsilon\frac{g_{3\alpha\beta}}{\chi_d}-V_{\alpha\beta}^2
\end{split}
\eeq
where $\epsilon$ is a positive constant.

From this equation we see if $g_{3\alpha\beta}<0$, it suppresses the interlayer pairing instability, while it enhances this instability if $g_{3\alpha\beta}>0$. This confirms the physical picture at the beginning of this appendix. But since $g_{3\alpha\beta}$ is assumed to be very small, the potential pairing instability will only occur at extremely low temperatures and hence we ignore it.

\section{Derivation of the formula for interlayer electric conductivity} \label{appkubo}

In this appendix we sketch the derivation of the formula for interlayer electric conductivity (\ref{masterformula}).

Starting from the Hamiltonian (\ref{multilayerHamiltonian}), we replace $t$ by $te^{ieAd}$ where $A$ is the external field. The partition function of the system in the presence of the external field is
\beq
\begin{split}
&\mc{Z}[A]\\
&\
=\int[Dc]e^{-\sum_\alpha S_\alpha+\int d\tau\sum_{i\alpha}t(e^{ieAd}c^\dag_{i,\alpha+1}c_{i,\alpha}+{\rm h.c.})}
\end{split}
\eeq
where $S_a$ is the Euclidean action of the $a$th layer and the second term in the exponent is the action corresponding to interlayer electron tunneling.

Suppose this external field induces current $j$, going to the frequency-momentum space, the conductivity in the linear response regime is
\beq \label{Kubo0}
\sigma(\omega)=\frac{1}{i\omega}\frac{j(\omega)}{A(\omega)}=\frac{1}{i\omega}\frac{\delta^2}{\delta A(\omega)^2}\ln\left(\frac{\mc{Z}[A]}{Z[0]}\right)
\eeq
Expanding $\ln\left(\frac{\mc{Z}[A]}{Z[0]}\right)$ in terms of $A$, the quadratic term is
\beq \label{quadraticinA}
\begin{split}
&-\frac{t(ed)^2}{2}\int d\tau A(\tau)^2\sum_{i\alpha}\la c_{i,\alpha+1}^\dag c_{i,\alpha}+{\rm h.c.}\ra
-\frac{(ted)^2}{2}\int d\tau\cdot\\
&\int d\tau' A(\tau)A(\tau')
\sum_{i\alpha,i'\alpha'}\Big\la\left(c^\dag_{i,\alpha+1}(\tau)c_{i,\alpha}(\tau)-{\rm h.c.}\right)\\
&\cdot\left(c_{i',\alpha'+1}^\dag(\tau')c_{i',\alpha'}(\tau')-{\rm h.c.}\right)\Big\ra
\end{split}
\eeq
Assuming different layers are decoupled and expanding to the leading order of $t$, the above expression in frequency-momentum space becomes
\beq
\begin{split}
&\frac{N(ted)^2}{\beta}\sum_{k\omega_1\omega_2}|A(i\omega_1)|^2\\
&\quad
\cdot\mc{G}(\vec k,i\omega_2)\left(\mc{G}(\vec k,i\omega_2)-\mc{G}(\vec k,i(\omega_1+\omega_2))\right)
\end{split}
\eeq
with the electron Green's function on each layer
\beq
\mc{G}(\vec k,i\omega)=\int d^2xd\tau\la T\{c_i(\tau)c_{0}(0)\}\ra e^{-i\vec k\cdot\vec r_i+i\omega\tau}
\eeq

By (\ref{Kubo0}),
\beq \label{Kubo1}
\sigma(\omega)=\frac{K(i\omega_n\rightarrow\omega+i\eta)}{i\omega}
\eeq
with
\beq
\begin{split}
&K(i\omega_n)=\frac{N(ted)^2}{\beta}\cdot\\
&\ \sum_{\vec k\omega_m}\mc{G}(\vec k,i\omega_m) \left(\mc{G}(\vec k,i\omega_n)-\mc{G}(\vec k,i(\omega_n+\omega_m))\right)
\end{split}
\eeq
Plugging spectral decomposition
\beq
\mc{G}(\vec k,i\omega)=\int d\Omega\frac{\mc{A}(\vec k,\Omega)}{i\omega-\Omega}
\eeq
into (\ref{Kubo1}) and taking its real part, we get the interlayer electric conductivity at frequency $\omega$:
\beq
\begin{split}
&\sigma(\omega)=-\pi N(ted)^2\cdot\\
&\sum_k\int d\Omega\mc{A}(\vec k,\Omega)\mc{A}(\vec k,\omega+\Omega)\frac{f(\omega+\Omega)-f(\Omega)}{\omega}
\end{split}
\eeq
Taking the limit $\omega\rightarrow 0$, we get (\ref{masterformula}) as the DC interlayer electric conductivity.

\section{Calculations of the imaginary parts of the spinon self-energies in the IFL and MFL regimes, and of the in-plane conductivity in the MFL regime} \label{appcalculations}

In this appendix we first provide more details of the calculations of the imaginary parts of the spinon self-energies in the IFL and MFL regimes, namely, (\ref{spinonselfenergyc}) and (\ref{spinonselfenergyb}), respetively. Then we also sketch the calculation of the in-plane conductivity in the MFL regime.

Both self-energies are calculated from an integral of the form
\beq
\begin{split}
&\Sigma_f(\vec k,i\omega)\\
&\quad
=v_{F}^2T\sum_{\Omega}\int_{\vec q}D(\vec q,i\Omega)\mc{G}_f^{(0)}(\vec k-\vec q,i(\omega-\Omega))
\end{split}
\eeq
Whether the result is for chemical potential tuned Mott transition or bandwidth controlled Mott transition depends on whether the gauge field propagator is of the form (\ref{rpachemical}) or (\ref{rpabandwidth}).

We first use spectral decomposition to write the above integral as
\beq
\begin{split}
&\Sigma_f(\vec k,i\omega)
=-\frac{v_F^2T}{\pi}\sum_{\Omega}\int_{\vec q}\\
&\quad
\cdot\int d\Omega_1d\Omega_2\frac{{\rm Im}D(\vec q,\Omega_1)\mc{A}_f^{(0)}(\vec k-\vec q,\Omega_2)}{(i\Omega-\Omega_1)(i(\omega-\Omega)-\Omega_2)}
\end{split}
\eeq
Now we can carry out the Mastubara frequency summation and get
\beq
\begin{split}
\Sigma_f(\vec k,i\omega)
=\frac{v_F^2}{\pi}\int d\Omega_1d\Omega_2[n(\Omega_1)+1-f(\Omega_2)]\\
\cdot\int_{\vec q}\frac{{\rm Im}D(\vec q,\Omega_1)\mc{A}_f^{(0)}(\vec k-\vec q,\Omega_2)}{i\omega-\Omega_1-\Omega_2}
\end{split}
\eeq
Performing the analytic continuation to real frequency $i\omega\rightarrow\omega+i\eta$ and taking the imaginary part, we get
\beq \label{singleparticlesr}
\begin{split}
\Sigma_f''(\vec k,\omega)=-v_F^2\int d\Omega[n(\Omega)+1-f(\omega-\Omega)]\\
\cdot\int_{\vec q}{\rm Im}D(\vec q,\Omega)\mc{A}_f^{(0)}(\vec k-\vec q,\omega-\Omega)
\end{split}
\eeq
Now we need to carry out the integral over $\vec q$. To do that, we need to specify the gauge field propagator.

\subsection{Calculation of the imaginary part of the spinon self-energy in the IFL regime}

In the IFL regime in the chemical potential tuned Mott transition, the gauge field propagator is given by (\ref{rpachemical}). As noted before, the important region of integral involves $q_\parallel\sim q_\perp^2\sim\Omega\sim\omega\ll q_\parallel^2$, so we can ignore the $q_\parallel$ dependence of the gauge field propagator. Now the integral over $q_\parallel$ can be carried out and we get
\beq \label{spinonselfenergycapp}
\begin{split}
&\Sigma_f''(\vec k,\omega)=\\
&\quad
-\frac{v_F}{4\pi^2}\int d\Omega\frac{\cosh\left(\frac{\beta\omega}{2}\right)} {\sinh\left(\beta\left(\Omega-\frac{\omega}{2}\right)\right)+\sinh\left(\frac{\beta\omega}{2}\right)}\\ &\quad\quad\qquad
\cdot\int dq_\perp\frac{k_0\frac{\Omega}{|q_\perp|}}{\left(k_0\frac{\Omega}{q_\perp}\right)^2+\left(\chi_dq_\perp^2+\rho_s\right)^2}
\end{split}
\eeq
The integral over $q_\perp$ is
\beq
\begin{split}
&\int dq_\perp\frac{k_0\frac{\Omega}{|q_\perp|}}{\left(k_0\frac{\Omega}{q_\perp}\right)^2+\left(\chi_dq_\perp^2+\rho_s\right)^2} =\\
&\qquad\qquad\qquad
\left\{
\begin{array}{lr}
\frac{2k_0\Omega}{\rho_s^2}\ln\left(\frac{\Omega_0}{a|\Omega|}\right), &|\Omega|\lesssim\Omega_c\\
\frac{2k_0\Omega}{\chi_d^{\frac{2}{3}}(k_0\Omega)^{\frac{4}{3}}}, &|\Omega|\gtrsim\Omega_c
\end{array}
\right.
\end{split}
\eeq
where the characteristic frequency scale $\Omega_c=\frac{a^{(c)}\rho_s^{\frac{3}{2}}}{k_0\sqrt{\chi_d}}$ with a constant $a^{(c)}$ that is on the order of unity. Plugging this result in (\ref{spinonselfenergycapp}), performing the integral over $\Omega$ and using that in the IFL regime $T\gg\Omega_c$, we get (\ref{spinonselfenergyc}).

\subsection{Calculation of the imaginary part of the spinon self-energy in the MFL regime}

In the MFL regime of the bandwidth controlled Mott transition, the gauge field propagator is given by (\ref{rpabandwidth}). Again, the important region of integral involves $q_\parallel\sim q_\perp^2\sim\Omega\sim\omega\ll q_\parallel^2$, so we can ignore the $q_\parallel$ dependence of the gauge field propagator. After carrying out the integral over $q_\parallel$ and $q_\perp$, we get
\beq
\begin{split}
&\Sigma_f''(\vec k,\omega)=\\
&\quad
-\frac{v_Fk_0}{4\pi^2\sigma_0^2\rho_s^2}\int d\Omega\frac{\cosh\left(\frac{\beta\omega}{2}\right)} {\sinh\left(\beta\left(\Omega-\frac{\omega}{2}\right)\right)+\sinh\left(\frac{\beta\omega}{2}\right)}\\ &\quad
\cdot\left[\ln\left(1+\left(\frac{\Omega_b}{\Omega}\right)^2\right) +\frac{\Omega_b}{|\Omega|}\left(\frac{\pi}{2}-\arctan\left(\frac{\Omega_b}{|\Omega|}\right)\right)\right]
\end{split}
\eeq
where the characteristic frequency scale $\Omega_b=\frac{a^{(b)}\sigma_0\rho_s^2}{k_0}$ with another constant $a^{(b)}$ on the order of unity. Finally, performing the integral over $\Omega$ and using that in the MFL regime $T\gg\Omega_b$, we get (\ref{spinonselfenergyb}).

\subsection{Calculation of the in-plane conductivity in the MFL regime} \label{appsubsectransportsrMFL}

In order to calculate the in-plane conductivity, we need to calculate the transport scattering rate of spinons, which is the imaginary part of the spinon self-energy multiplied by an additional factor $q_\perp^2/k_0^2$. So from (\ref{singleparticlesr}), we see the transport scattering rate of spinons is
\beq
\begin{split}
\gamma_{\rm tr}
=&\frac{v_F^2}{k_0^2}\int d\Omega[n(\Omega)+f(\Omega)]\\
&\quad
\cdot\int_{\vec q}q_\perp^2{\rm Im}D(\vec q,\Omega)\mc{A}_f^{(0)}(\vec k-\vec q,\omega-\Omega)
\end{split}
\eeq

As before, because the important region of integral involves $q_\parallel\sim q_\perp^2\Omega\sim\omega\ll q_\parallel^2$, we will ignore the $q_\parallel$ dependence of the gauge field propagator (\ref{rpabandwidth}). Then the integral can be done and we find the dominant contribution is
\beq
\gamma_{\rm tr}\sim T^2
\eeq
This implies the in-plane conductivity behaves as
\beq
T^{-2}
\eeq

\section{Validity of the perturbative calculations of the interlayer electric conductivity} \label{appvalidity}

In this appendix we discuss the validity of the perturbative calculation of the interlayer electric conductivity, which only keeps the leading order terms in $t$, the interlayer electron tunneling amplitude. Notice in general one needs to consider all types of interlayer interactions discussed in the main text, but here, for simplicity, we only consider interlayer electron tunneling specified by the model (\ref{multilayerHamiltonian}).

According to (\ref{quadraticinA}), schematically, each higher order term in the perturbative expansion picks up one more $t^2|\la b\ra|^4\mc{G}_f^2$ factor compared to the previous term. This implies, due to the structure of (\ref{spinonspectral}), the $n$th term in the perturbative expansion of $\sigma$ has the structure
\beq
\frac{t^{2n}|\la b\ra|^{4n}}{\Sigma_f''^{2n-1}}
\eeq
So the ratio of the $n+1$th term to the $n$th term can be schematically written as
\beq \label{ratio}
\frac{t^2|\la b\ra|^4}{\Sigma_f''^2}
\eeq

For the IFL regime of the chemical potential tuned Mott transition, we expect (\ref{ratio}) is of the order of $t^2|\la b\ra|^4/T^{\frac{4}{3}}\sim t^2(g-g_c)^{4\beta}/T^{\frac{4}{3}}$. Because $T\gg\rho_s^{\frac{3}{2}}\sim(g-g_c)^{3\nu}$ in the IFL regime, we have $t^2(g-g_c)^{4\beta}/T^{\frac{4}{3}}\ll t^2(g-g_c)^{4(\beta-\nu)}$. In the universality class of 2D dilute bose gas $\beta=\nu=\frac{1}{2}$, so this ratio is much smaller than unity as long as the system is close to the quantum critical point and $t$ is small, which means the higher order terms in the perturbative expansion is much smaller than the leading order contribution. Moreover, as shown in Ref. \onlinecite{SenthilLee2009}, the in-plane electric conductivity scales as $T^{-\frac{4}{3}}$, so the interlayer electric conductivity in (\ref{sigmaifl}) is much smaller than the in-plane one as long as the temperature is low enough. Therefore, it is expected that the perturbative series in the IFL regime of the chemical potential tuned Mott transition will converge.

The issue is a little more subtle in the MFL regime of the bandwidth controlled Mott transition, which lies in the temperature regime $(g-g_c)^{\frac{4\beta}{1+\eta}}\ll T\ll(g-g_c)^{\frac{2\beta}{1+\eta}}$, where we have used $\rho_s\sim(g-g_c)^\nu$ and the scaling relation $\beta=\frac{\nu(1+\eta)}{2}$ in the universality class of 3D XY model. (\ref{ratio}) in this case is expected to be of the order of $t^2|\la b\ra|^4/\left(T\ln\frac{T}{\rho_s^2}\right)^2$. Ignoring the logarithmic correction, the condition under which the perturbative calculation is valid is that $T\gtrsim t|\la b\ra|^2\sim t(g-g_c)^{2\beta}$. So long as the interlayer electron tunneling amplitude $t$ is very small, we can still have $(g-g_c)^{\frac{4\beta}{1+\eta}}>t(g-g_c)^{2\beta}$, then the higher order terms in the perturbative series are small compared to the leading order one. Moreover, according to Appendix \ref{appsubsectransportsrMFL}, the in-plane conductivity in this regime scales as $T^{-2}$, which is much larger than the leading order contribution to the interlayer conductivity in the low energy regime. Therefore, the perturbative calculation is expected to be valid in the regime where $g-g_c\gtrsim t^{\frac{1+\eta}{2\beta(1-\eta)}}\approx t^{\frac{1}{2\beta}}$.

On the other hand, if the parameters of the system are in a narrow window where $g-g_c$ is extremely small so that $g-g_c\lesssim t^{\frac{1+\eta}{2\beta(1-\eta)}}$ and the temperature is so low that $T\lesssim t(g-g_c)^{2\beta}$, the higher order terms in the perturbative series can be larger than the leading order term, so the perturbative calculation breaks down. In this case, we expect the temperature dependence will be more singular than $T^{-1}$ but no more singular than $T^{-2}$, which is the Fermi liquid form. But as long as the interlayer electron tunneling amplitude $t$ is small, this only occurs in an extremely narrow window of the parameter space, so it may be more difficult to detect experimentally.

\end{appendix}

\bibliography{dimdcp}


\end{document}